\documentclass[final,5p,times,twocolumn]{elsarticle}

\usepackage{amssymb}
\usepackage{amsmath}

\usepackage{mathtools}

\usepackage{verbatim}

\usepackage{graphicx} 
\usepackage{subcaption} 

\usepackage{color}
\definecolor{amethyst}{rgb}{0.6, 0.4, 0.8}
\definecolor{alizarin}{rgb}{0.82, 0.1, 0.26}
\definecolor{green}{rgb}{0.55, 0.71, 0.0}
\definecolor{apricot}{rgb}{0.98, 0.81, 0.69}
\definecolor{auburn}{rgb}{0.43, 0.21, 0.1}
\definecolor{babyblueeyes}{rgb}{0.63, 0.79, 0.95}
\definecolor{bittersweet}{rgb}{1.0, 0.44, 0.37}
\definecolor{arsenic}{rgb}{0.23, 0.27, 0.29}
\usepackage{ulem}
\usepackage{soul}
\newcommand{\cfsout}{\bgroup\markoverwith{\textcolor{red}{\rule[0.5ex]{2pt}{0.4pt}}}\ULon}

\providecommand{\ro}[1]{\mathrm{#1}}
\providecommand{\ca}[1]{\mathcal{#1}}

\usepackage[colorlinks=true,
            linkcolor=blue,
            urlcolor=blue,
            citecolor=blue]{hyperref}

\usepackage{lineno}

\journal{Physics Dark Universe}

\begin{document}
\begin{frontmatter}



\title{Recasting Experimental Constraints on Relativistic Magnetic Monopoles} 

\author[a,b,c,d]{Daniele Perri}\ead{dperri@sissa.it}
\author[e]{Michele Doro}\ead{michele.doro@unipd.it}
\author[a,b,c,f,g]{Takeshi Kobayashi}\ead{takeshi.kobayashi@sissa.it}

\affiliation[a]{
  organization={SISSA, International School for Advanced Studies},
  postcode={via Bonomea 265, 34136},
  city={Trieste},
  country={Italy},
}

\affiliation[b]{
  organization={INFN, Sezione di Trieste},
  postcode={via Valerio 2, 34127},
  city={Trieste},
  country={Italy},
}

\affiliation[c]{
  organization={IFPU, Institute for Fundamental Physics of the Universe},
  postcode={via Beirut 2, 34014},
  city={Trieste},
  country={Italy},
}

\affiliation[d]{
  organization={Institute of Theoretical Physics, Faculty of Physics, University of Warsaw},
  postcode={ul. Pasteura 5, PL-02-093},
  city={Warsaw},
  country={Poland},
}

\affiliation[e]{
  organization={University of Padova, Department of Physics and Astronomy},
  postcode={I-35131},
  city={Padova},
  country={Italy},
}

\affiliation[f]{
  organization={Kobayashi-Maskawa Institute for the Origin of Particles and the Universe (KMI), Nagoya University},
  city={Nagoya},
  postcode={464-8602},
  country={Japan},
}

\affiliation[g]{
  organization={Kavli Institute for the Physics and Mathematics of the Universe (WPI), 
University of Tokyo},
  city={Kashiwa},
  postcode={277-8583},
  country={Japan},
}

\begin{abstract}
Magnetic monopoles with masses up to $10^{14}$ GeV can be accelerated to relativistic velocities in Galactic and intergalactic magnetic fields. The cosmic flux of relativistic monopoles is constrained by various experiments, with the limits given as functions of the monopole velocity (Lorentz factor) at the detectors. The velocity, however, is usually treated as a free parameter due to the ambiguity in the computation of the acceleration before the monopoles arrive at Earth.
We explicitly evaluate the velocity by exploiting recent studies on cosmic magnetic fields and the monopole acceleration therein, to recast experimental limits in terms of the mass of monopoles. By applying our method to various terrestrial experiments, including 
the Pierre Auger Observatory, IceCube, MACRO, and the upcoming Cherenkov Telescope Array Observatory, 
as well as to astrophysical constraints, we report limits on the flux of monopoles for a wide range of monopole masses. 
We also highlight the role of monopoles as messengers of cosmic magnetic fields, and discuss the possibility of using monopole experiments to probe intergalactic magnetic fields. 
\end{abstract}




\end{frontmatter}



\section{Introduction}
\label{sec:intro}

In any unified gauge theory in which the electromagnetic U(1) is embedded in a spontaneously broken semisimple group, magnetic monopoles (MMs) exist as soliton solutions~\cite{tHooft:1974kcl,Polyakov:1974ek}. 
MMs are then inevitably produced in the early universe during a symmetry breaking phase transition~\cite{Preskill:1979zi,Zeldovich:1978wj}. 
Moreover, they can be produced thermally~\cite{Turner:1982kh}, or through a Schwinger production in primordial magnetic fields~\cite{Kobayashi:2021des}. 

While Dirac's quantization condition~\cite{Dirac:1931kp} requires the magnetic charge of MMs to be integer multiples of $2 \pi / e$, 
the mass can vary by many orders of magnitude, depending on the symmetry breaking scale of the unified theory. 
Here, the intermediate to low mass range\footnote{For recent explicit constructions of light MMs, see, e.g., \cite{Lazarides:2024niy,Kephart:2025tik}.} (i.e. $m \lesssim 10^{14}\, \ro{GeV}$) is particularly interesting from the astrophysical point of view, since MMs with such masses can be accelerated to relativistic velocities 
by cosmic magnetic fields~\cite{Wick:2000yc,Perri:2023ncd}.
Relativistic MMs leave various observable signals in direct and indirect detection, 
and thus have been a subject of extensive experimental searches. 
(See e.g. \cite{Giacomelli:2000de,Balestra:2011ks,Mavromatos:2020gwk} for reviews.)
These have placed various upper limits on the cosmic flux of relativistic MMs, which, however, strongly depend on the Lorentz factor of the MMs arriving at the detectors of the individual experiments. 
Mapping the limits onto the fundamental properties of MMs thus requires an accurate understanding of the MM velocity,
which had been a challenging task due to the modeling ambiguities in the 
MM acceleration in the cosmic space.

Recently in \cite{Perri:2023ncd}, we carried out a detailed analysis of MM acceleration in the magnetic fields in the Milky Way Galaxy and in the intergalactic space, by taking into account the backreaction of the MMs on the fields. 
Using the result, in this work 
we translate experimental limits on the MM flux into functions of the MM mass. 
By evaluating the acceleration of MMs in the cosmic space, and the deceleration in Earth, we recast limits from
MACRO~\cite{MACRO:2002kki,MACRO:2002jdv}, IceCube~\cite{IceCube:2012, IceCube:2014xnp, IceCube:2015agw,IceCube:2021eye}, Pierre Auger Observatory~\cite{PierreAuger:2016imq}, and the upcoming Cherenkov Telescope Array Observatory~\cite{CTAConsortium:2017dvg}. 
Combining these with astrophysical constraints
such as the Parker bound~\cite{Parker:1970xv,Turner:1982ag}, 
we present a comprehensive study of flux limits for intermediate to low mass MMs. 

We also highlight the role of MMs as messengers of cosmic magnetic fields. 
The intermediate to low mass MMs do not cluster with the Milky Way, and hence they are accelerated in both the Galactic magnetic fields (GMFs) and intergalactic magnetic fields (IGMFs). 
Here, while the basic properties of GMFs are determined~\cite{Widrow:2002ud,Haverkorn:2014jka,Unger:2023lob}, IGMFs still have large uncertainties in both their strength and coherence length~\cite{Durrer:2013pga,AlvesBatista:2021sln,Neronov:2021xua}. 
We clarify the conditions for MM acceleration in IGMFs to be dominant over that in GMFs. Then we show that some of the existing MM experiments already have enough sensitivity to test the acceleration in IGMFs, and thereby have the potential to probe IGMFs. 

This paper is organized as follows.   
In \autoref{sec:MonAcc} we review the acceleration of MMs in cosmic magnetic fields. 
In \autoref{sec:results} we recast various experimental and astrophysical MM limits in terms of the MM mass. 
In \autoref{sec:concl} we summarize our results and conclude. 
Technical calculations for some of the experiments are relegated to the appendices. 

Throughout this paper we use Heaviside--Lorentz units, with $c = \hbar = k_B = 1$. 
We hence use $v$ and $\beta = v / c$ interchangeably to denote the velocity of MMs. 
We denote the MM mass by~$m$, the amplitude of the magnetic charge by~$g$, and the Dirac charge by $g_{\ro{D}} \equiv 2 \pi / e \approx 21$.

\section{Monopole acceleration in Cosmic Magnetic Fields}
\label{sec:MonAcc}

The acceleration of MMs in GMFs and IGMFs was evaluated in our previous work~\cite{Perri:2023ncd}. There we modeled the MM dynamics as a random walk through cells of uniform magnetic field, while setting the size of the individual cells to the field's coherence length. 
We start by briefly reviewing the results.\footnote{The random walk through the magnetic field cells gives rise to a variance of the MMs' velocity, which however is tiny as long as the total MM number is much greater than unity. 
(See Appendix of \cite{Kobayashi:2023ryr}; $v_{\ro{I}}$ and $v_{\ro{G}}$ we introduce below are derived using the second line of their Eq.~(A15).)
We thus focus on the MMs' mean velocity in this paper.\label{fn:2}}

\subsection{Acceleration in intergalactic magnetic fields}

\begin{figure}[t!]
    \centering
    \includegraphics[width=\linewidth]{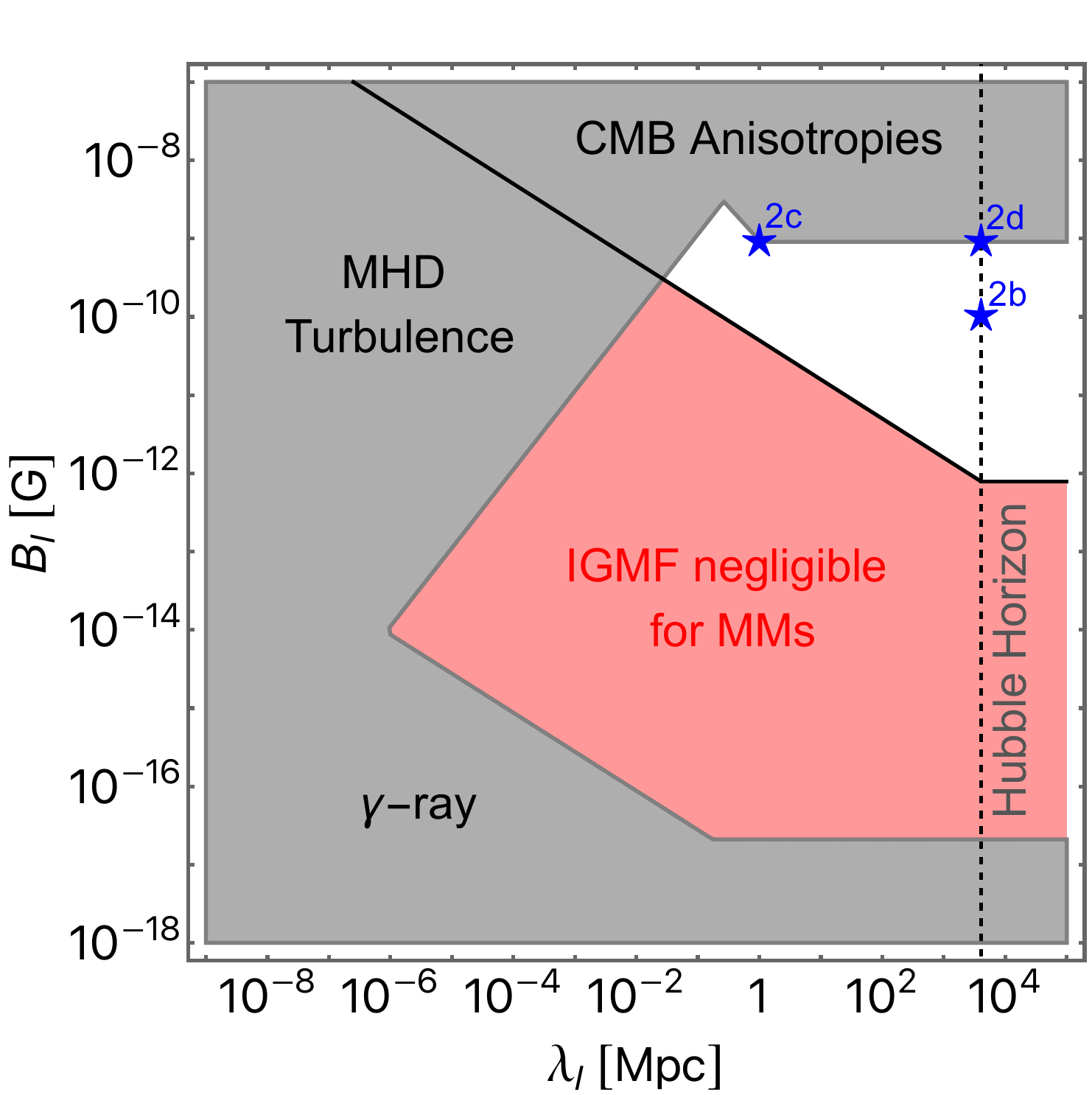}
    \caption{Parameter space of IGMF strength and coherence length. 
The gray region is excluded by various constraints~\cite{Durrer:2013pga}.
The red region shows where IGMFs have negligible effects on the MM velocity at Earth. The blue stars indicate benchmark values that are used in \autoref{fig:gammabetam}.
}
    \label{fig:igmf_limits}
\end{figure}

IGMFs in cosmic voids are constrained by a number of experiments (see \cite{AlvesBatista:2021sln,Neronov:2021xua} for recent reviews).
On one hand, gamma-ray observations \cite{Tavecchio:2010mk,Neronov2010,Ackermann2018,MAGIC:2022piy,HESS:2023zwb, Blunier:2025ddu} suggest the existence of IGMFs based on the non-observation of cascade emissions that are expected to originate from distant blazars. 
This provides a lower limit of 
$B_{\ro{I}} \gtrsim 10^{-17}\, \ro{G}$ if the coherence length~$\lambda_{\ro{I}}$ is of Mpc scale or larger; otherwise the lower limit further improves as~$\lambda_{\ro{I}}^{-1/2}$.
On the other hand, CMB observations \cite{Barrow:1997,Planck:2015zrl,Jedamzik:2018itu} set an upper limit of $B_{\ro{I}} \lesssim 10^{-9}\, \ro{G}$.\footnote{There have also been recent claims 
\cite{Pavicevic:2025gqi,Pignataro:2025ntd,Jedamzik:2025cax}
hinting IGMFs with strengths not too far from the CMB upper limit.}
Further constraints are set by discussions on
magnetohydrodynamics turbulence~\cite{Grasso:2000wj,Durrer:2013pga}. The combinations of $B_{\ro{I}}$ and $\lambda_{\ro{I}}$ excluded by these constraints are illustrated in \autoref{fig:igmf_limits} as the gray shaded region. 

If the coherence length is larger than the Hubble radius, i.e., $\lambda_{\ro{I}} > 1 / H_0$ where 
$H_0 \approx 1.4 \times 10^{-33}\, \ro{eV}$, the IGMF is effectively homogeneous.
After being accelerated in a homogeneous IGMF for a Hubble time,\footnote{Even if the IGMFs have a primordial origin and the MMs have been accelerated from high redshifts, the present-day MM velocity is unaffected at the order-of-magnitude level \cite{Perri:2023ncd}.} 
the product of the MMs' velocity and Lorentz factor $\gamma = (1-v^2)^{-1/2}$ 
becomes:
\begin{equation}
 (\gamma v )_0 \sim 
\frac{ g B_{\ro{I}}}{ m H_0}.
\label{eq:2.2}
\end{equation}
On the other hand for an inhomogeneous field, i.e., $\lambda_{\ro{I}} < 1 / H_0$, the average MM velocity after a Hubble time is written as,
\begin{equation}
\left( \gamma v \right)_0 \sim
  \begin{dcases}
\frac{gB_{\mathrm{I}}}{m H_0}
  & \mathrm{for}\, \, \,  
m > \frac{g B_{\mathrm{I}}}{\lambda_{\mathrm{I}} H_0^2},
 \\
\left( \frac{gB_{\mathrm{I}} \lambda_{\mathrm{I}}}{m } \right)^{2/3}
\frac{1}{(\lambda_{I} H_0 )^{1/3}}
  & \mathrm{for}\, \, \,  
\frac{g B_{\mathrm{I}}  \lambda_{\mathrm{I}}^{1/2}}{H_0^{1/2}} < m < 
\frac{g B_{\mathrm{I}}}{\lambda_{\mathrm{I}} H_0^2} ,
 \\
 \frac{gB_{\mathrm{I}} \lambda_{I}}{m }
\frac{1}{(\lambda_{\mathrm{I}} H_0)^{1/2}}
  & \mathrm{for}\, \, \,  
m < \frac{g B_{\mathrm{I}} \lambda_{\mathrm{I}}^{1/2}}{H_0^{1/2}}.
 \end{dcases}
\label{eq:nonhomoGen}
\end{equation}
The first line applies to MMs that are sufficiently heavy such that they each travel a distance smaller than~$\lambda_\ro{I}$.
In the second and third lines, the MMs travel distances larger than~$\lambda_{\ro{I}}$, with the final velocity being nonrelativistic and relativistic, respectively.

Depending on their number density, the MMs can drain the IGMF energy before being accelerated to~$v_0$.
This triggers an energy oscillation between the MMs and IGMFs (a magnetic analog of the Langmuir oscillation), 
and the oscillation-averaged MM velocity is of order:
\begin{equation}
\label{eq:gamma_v_max}
(\gamma v)_{\ro{max}} \simeq
  \begin{dcases}
\frac{B_{\mathrm{I}}^2}{4 \pi m F_{\ro{I}}}
  & \mathrm{for}\, \, \,  
B_{\mathrm{I}}^2 \ll 8 \pi m F_{\ro{I}},
 \\
\frac{B_{\mathrm{I}}^2}{8 \pi m F_{\ro{I}}}
  & \mathrm{for}\, \, \,  
B_{\mathrm{I}}^2 \gg 8 \pi m F_{\ro{I}}.
 \end{dcases}
\end{equation}
Here $F_{\ro{I}}$ is the intergalactic MM flux per area per time per solid angle, in the CMB rest frame.
The first and second lines are solutions in the nonrelativistic and relativistic limits, respectively.

Using the expressions above, the MM velocity in the CMB rest frame induced by IGMFs can be written as
\begin{equation}
     v_{\ro{I}} = \ro{min.} \left\{ v_0,  v_{\mathrm{max}} \right\} .
\label{eq:v_MW}
\end{equation}

\subsection{Acceleration in Galactic magnetic fields}

Observations indicate that the Milky Way hosts GMFs
with an average amplitude $B_{\rm G} \sim 10^{-6} \, \mathrm{G}$ and coherence length $\lambda_{\mathrm{G}} \sim 1 \, \mathrm{kpc}$, 
within a magnetic region of size $R \sim 10\, \ro{kpc}$~\cite{Widrow:2002ud,Haverkorn:2014jka,Unger:2023lob}.
For MMs that originate from outside the Milky Way and propagate to Earth, the typical kinetic energy that they obtain from the GMFs is of 
\begin{equation}
\label{eq:ek}
m \left( \gamma_{\rm G} - 1 \right) \sim g B_{\ro{G}} \sqrt{R \lambda_{\ro{G}} } 
\sim 10^{11}\, \mathrm{GeV} \left(\frac{g}{g_{\mathrm{D}}} \right).
\end{equation}
There can also be MMs that are clustered with the Milky Way, 
however MMs with a single Dirac charge can stay clustered until today only if they have a mass  $ m \gtrsim 10^{18}\, \ro{GeV}$~\cite{Turner:1982ag,Kobayashi:2023ryr}.

A large flux of MMs in the Milky Way would dissipate the GMFs, instead of inducing a MM-magnetic field oscillation as in the case of IGMFs. This is because unclustered MMs pass through the Milky Way and hence do not completely return the energy to the GMFs, and for clustered MMs the oscillation rapidly evaporates due to Landau damping. 
Requiring the survival of GMFs imposes the so-called Galactic Parker bound on the MM flux~\cite{Parker:1970xv,Turner:1982ag, Parker:1987}, which will be presented in \autoref{sec:results}. 

\begin{figure*}[t!]
     \centering
     \begin{subfigure}[b]{0.49\linewidth}
         \centering
         \includegraphics[width=0.9\linewidth]{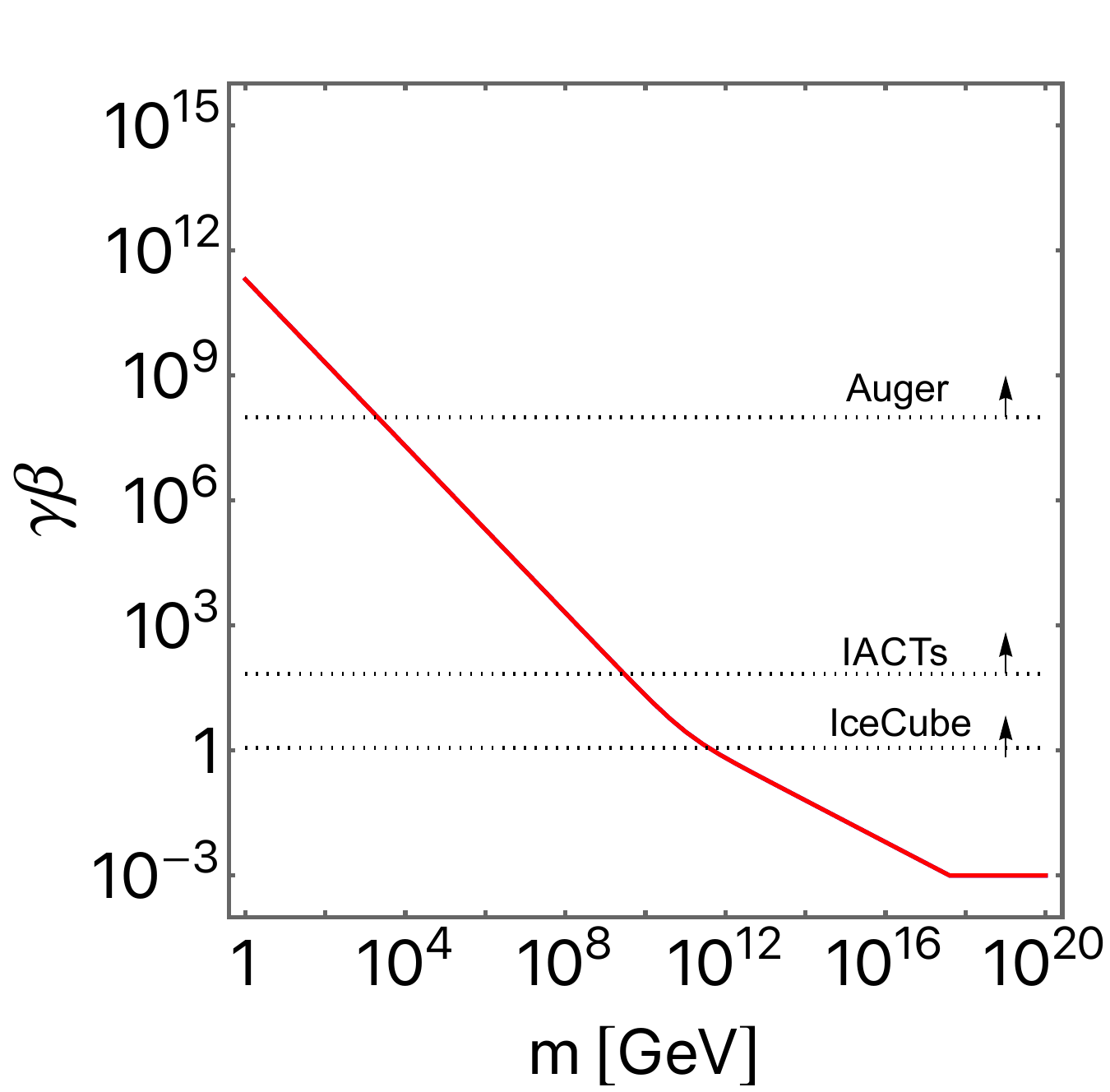}
         \caption{$B_{\rm I} \lesssim 10^{-12}\, G \left[ \min{\{1, \lambda_{\rm I} H_0\}} \right]^{-1/2}$.}
         \label{fig:vEarth11}
     \end{subfigure}
     \hfill
     \begin{subfigure}[b]{0.49\linewidth}
         \centering
         \includegraphics[width=0.9\linewidth]{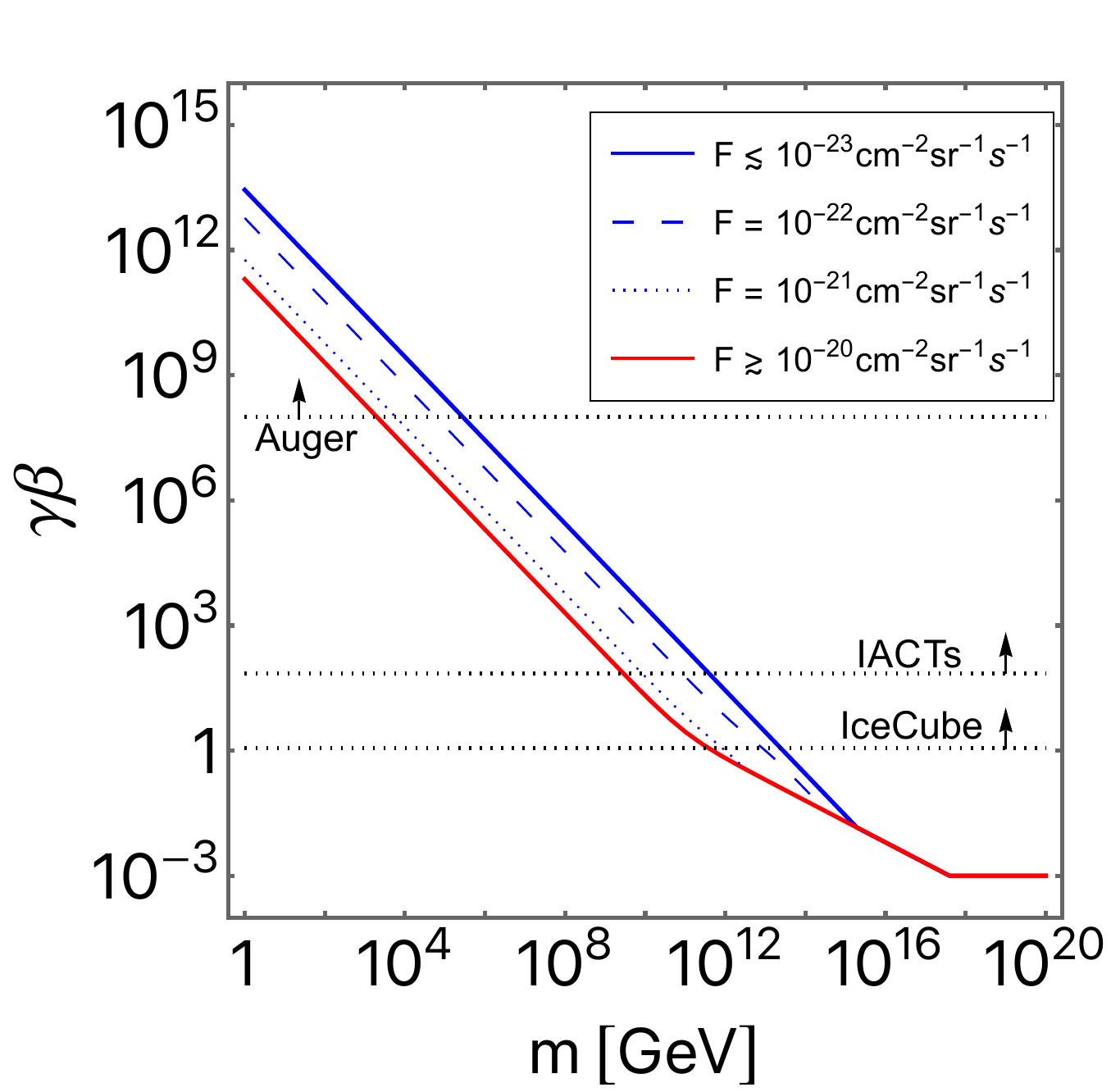}
         \caption{$B_{\rm I} = 10^{-10}\, G$,  $\lambda_{\rm I} \gtrsim 1/H_0$.}
         \label{fig:vEarth10}
     \end{subfigure}
     \hfill
     \begin{subfigure}[b]{0.49\linewidth}
         \centering
         \includegraphics[width=0.9\linewidth]{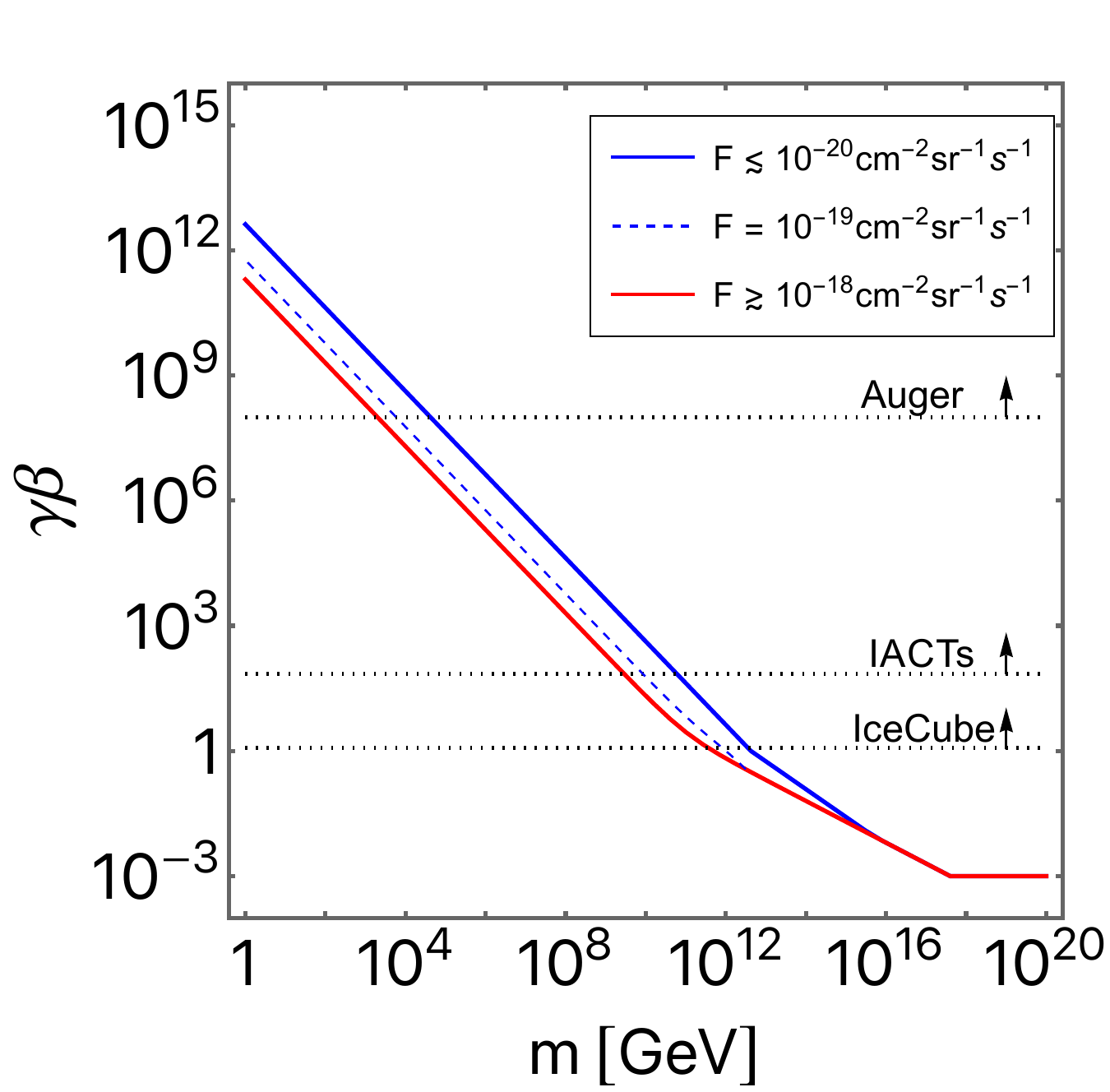}
         \caption{$B_{\rm I} = 10^{-9}\, G$, $\lambda_{\rm I} = 1\, \mathrm{Mpc}$.}
         \label{fig:vEarth9_1}
     \end{subfigure}
     \hfill
          \begin{subfigure}[b]{0.49\linewidth}
         \centering
         \includegraphics[width=0.9\linewidth]{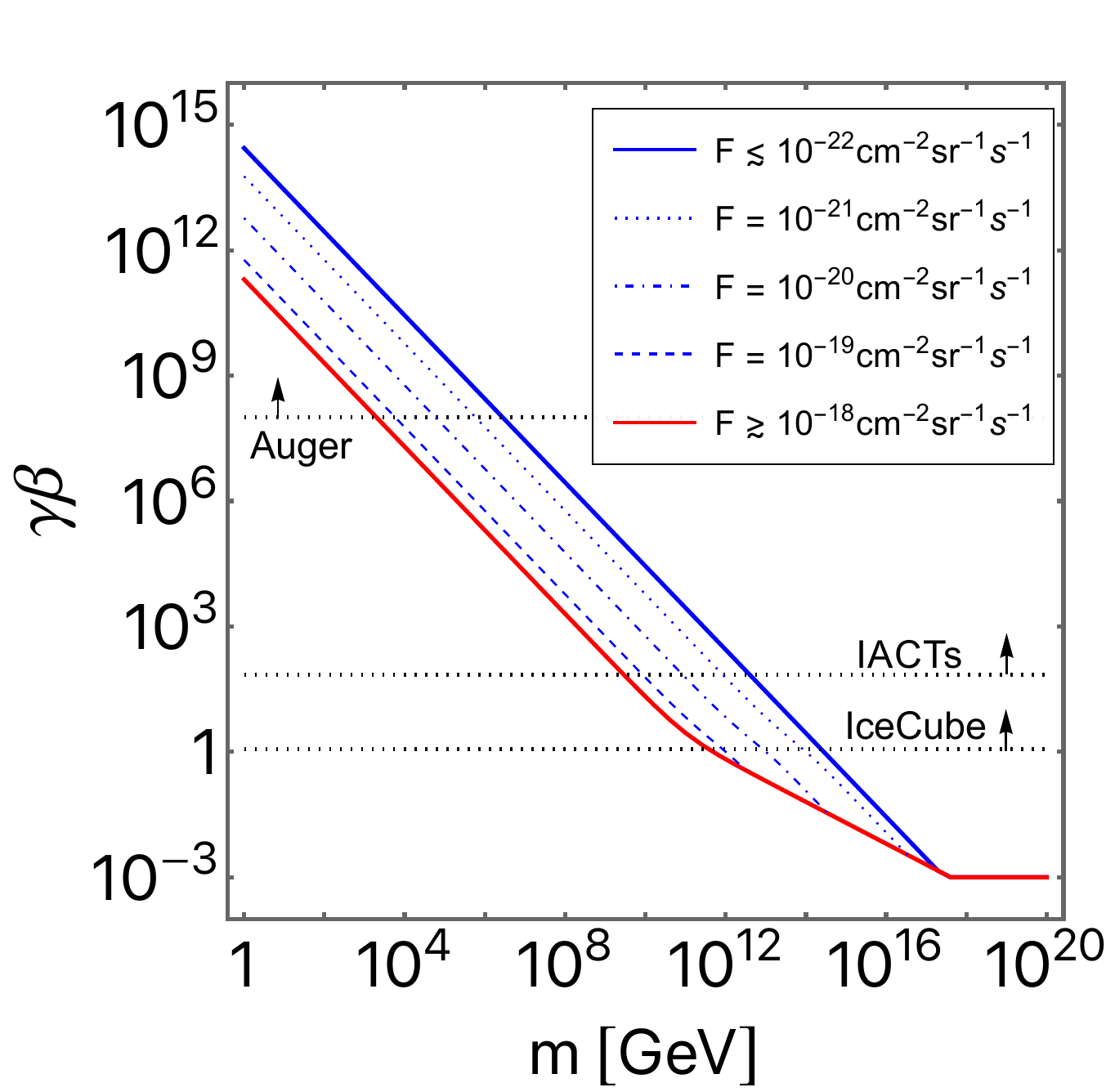}
         \caption{$B_{\rm I} = 10^{-9}\, G$,  $\lambda_{\rm I} \gtrsim 1/H_0$.}
         \label{fig:vEarth9}
     \end{subfigure}
     \caption{\label{fig:gammabetam}Product of the velocity and Lorentz factor of MMs at Earth, as a function of mass.
The MM charge is fixed to $g = g_{\mathrm{D}}$.
The IGMF strength and coherence length are varied in each panel.
In (a) the IGMF is too weak to affect the MM velocity.
In (b)--(d) the IGMFs can predominantly accelerate the MMs, depending on the MM flux. 
The horizontal black dotted lines indicate the sensitivity thresholds for IceCube, IACTs, and Auger.}        
\end{figure*}

\subsection{Monopole velocity at Earth}

MMs propagating to Earth from outside the Milky Way have been accelerated by the IGMFs and GMFs, which respectively source a velocity~$v_{\ro{I}}$ in the CMB rest frame (\autoref{eq:v_MW}), 
and $v_{\ro{G}}$ in the Milky Way's rest frame (\autoref{eq:ek}).
Notice that the Milky Way itself has a peculiar velocity of
$v_{\ro{p}} \sim 10^{-3}$ with respect to the CMB rest frame \cite{Kogut:1993ag}.\footnote{The MMs' gravitational infall velocity into the Milky Way, as well as the velocity of the Earth with respect to the Milky Way, are also of $10^{-3}$.}
Hence the MM velocity measured on Earth is written as\footnote{Energy losses of MMs from radiative emissions, and interactions with the interstellar and intergalactic media, 
are negligible for a wide range of parameters \cite{Kobayashi:2023ryr,Perri:2023ncd}. 
Earth's magnetic field is also negligible, since even if one assumes the MMs to be accelerated by a field $\sim 1 \, \ro{G}$ over the Earth radius $\sim 10^4\, \ro{km}$, the energy gain is $ 10^{-7}$ smaller than that from GMFs (cf. \autoref{eq:ek}).}
\begin{equation}
\label{eq:vEarth}
    v_{\ro{E}} = \ro{max.} \left\{ v_{\ro {I}}, v_{\ro{G}}, v_{\mathrm{p}} \right\} .
\end{equation}
If in particular $v_{\ro{G}}, v_{\rm p} < v_{\ro{max}} < v_0$, one obtains $v_{\ro{E}} = v_{\ro{max}}$, which is a function of the MM flux~$F_{\ro{I}}$ in the CMB rest frame (\autoref{eq:gamma_v_max}).
However in this case the MM flux in the CMB rest frame matches with that in the Earth's rest frame (for a detailed discussion on this point, 
we refer the reader to~\cite{Perri:2023ncd}). 
Hence for \autoref{eq:vEarth}, the flux variable can be identified with the MM flux measured on Earth.
In the rest of the paper, unless otherwise stated,
we use $F$ to denote the MM flux on Earth.\footnote{($v_{\ro{I}}$, $F_{\ro{I}}$, $F$) are equivalent to ($v_{\ro{CMB}}$, $F_{\ro{CMB}}$, $F_{\ro{E}}$) in \cite{Perri:2023ncd}.}

In order for the IGMFs to have a larger contribution to the MM velocity than GMFs, i.e. $v_{\ro{I}} > v_{\ro{G}}$,
one sees from \autoref{eq:v_MW} that both 
$v_0 > v_{\ro{G}}$ and $v_{\ro{max}} > v_{\ro{G}}$ need to be satisfied. 
In the small mass (thus relativistic) limit, the former condition can be rewritten by combining
\autoref{eq:2.2}, \autoref{eq:nonhomoGen} (third line), and \autoref{eq:ek} as
\begin{equation}
B_{\ro{I}} \gtrsim B_{\ro{G}} 
\left[
\frac{R \lambda_{\ro{G}} H_0^2}{\min \{ 1,  \lambda_{\ro{I}} H_0 \}}
\right]^{1/2}
\sim
\frac{ 10^{-12}\, \ro{G} }{ \left[
\min \{ 1,  \lambda_{\ro{I}} H_0 \}
\right]^{1/2} },
 \label{eq:BI_limit}
\end{equation}
while the latter is rewritten using 
\autoref{eq:gamma_v_max} (second line) and \autoref{eq:ek} as
\begin{equation}
 F \lesssim \frac{B_{\ro{I}}^2}{8 \pi g B_{\ro{G}} (R \lambda_{\ro{G}})^{1/2}}
\sim \frac{10^{-19}}{\, \mathrm{cm}^{2} \,\mathrm{s} \,\mathrm{sr}} \;
\left( \frac{g}{g_{\mathrm{D}}} \right)^{-1}
\left( \frac{B_{\ro{I}}}{ 10^{-9} \, \ro{G} } \right)^{2}.
\label{eq:FII}
\end{equation}
When both of these conditions are satisfied, the acceleration in IGMFs dominates over that in GMFs for low mass MMs.
In the $\lambda_{\ro{I}}$ - $B_{\ro{I}}$ plane in \autoref{fig:igmf_limits}, the black solid line shows where the condition in \autoref{eq:BI_limit} is saturated. 
In the red region below this line, IGMFs have negligible effects on the MM velocity at Earth, independently of~$F$. 
In particular, IGMFs with $B_{\ro{I}} \lesssim 10^{-12}\, \ro{G}$ are negligible for any $\lambda_{\ro{I}}$.
On the other hand, the condition in \autoref{eq:FII}
combined with the CMB limit $B_{\ro{I}} \lesssim 10^{-9}\, \ro{G}$ implies that terrestrial experiments have the potential to probe IGMF acceleration of MMs with a Dirac charge, if their sensitivity reaches down to $F \lesssim 10^{-19}\, \mathrm{cm}^{-2} \mathrm{s}^{-1} \mathrm{sr}^{-1} $.

In \autoref{fig:gammabetam} we show the MM velocity at Earth in \autoref{eq:vEarth}, 
as a function of mass, with the MM charge fixed to $g = g_{\ro{D}}$.
The IGMF strength and coherence length are varied in each panel.
Panel~\ref{fig:vEarth11} is for IGMFs 
below the threshold in \autoref{eq:BI_limit}
(i.e. within the red region in \autoref{fig:igmf_limits}), for which the MM velocity is determined solely by GMFs.
One sees that GMFs accelerate
MMs with $m \lesssim 10^{12}\, \ro{GeV}$
to relativistic velocities,
while having negligible effects for MMs with $m \gtrsim 10^{18}\, \ro{GeV}$, which thus travel with the 
peculiar velocity of the Milky Way, $v_{\mathrm{p}} \sim 10^{-3}$.

Panels~\ref{fig:vEarth10} to \ref{fig:vEarth9} are for IGMFs above the threshold.
The chosen sets of values for $\lambda_{\ro{I}}$ and $B_{\ro{I}}$
are also indicated in \autoref{fig:igmf_limits} by the blue stars.
Here, the IGMF contribution to the MM velocity is still subdominant if the MM flux exceeds the threshold in Eq.~(\ref{eq:FII}), in which case the velocity follows the same red line as in Panel~\ref{fig:vEarth11}.
On the other hand with smaller fluxes,
the IGMF contribution dominates unless the MMs are superheavy.
The IGMF-induced velocities are shown by the blue lines with different line styles, corresponding to different flux values. 
MMs with fluxes just below the threshold 
significantly backreact to the IGMFs such that their velocity is given by~$v_{\ro{max}}$,
while with much smaller fluxes the backreaction is negligible and the velocity approaches~$v_{0}$ which is independent of the flux value
(cf. discussions around Eq.~(\ref{eq:v_MW})).
The latter case is denoted by the blue solid line, 
which shows that even MMs as heavy as 
$m \sim 10^{14} \, \mathrm{GeV}$ can be accelerated to relativistic velocities in IGMFs.
The horizontal dotted lines in the plots indicate the sensitivity thresholds for the terrestrial experiments that we will discuss in the next section.

\begin{figure}[t!]
     \centering
     \includegraphics[width=0.95\linewidth]{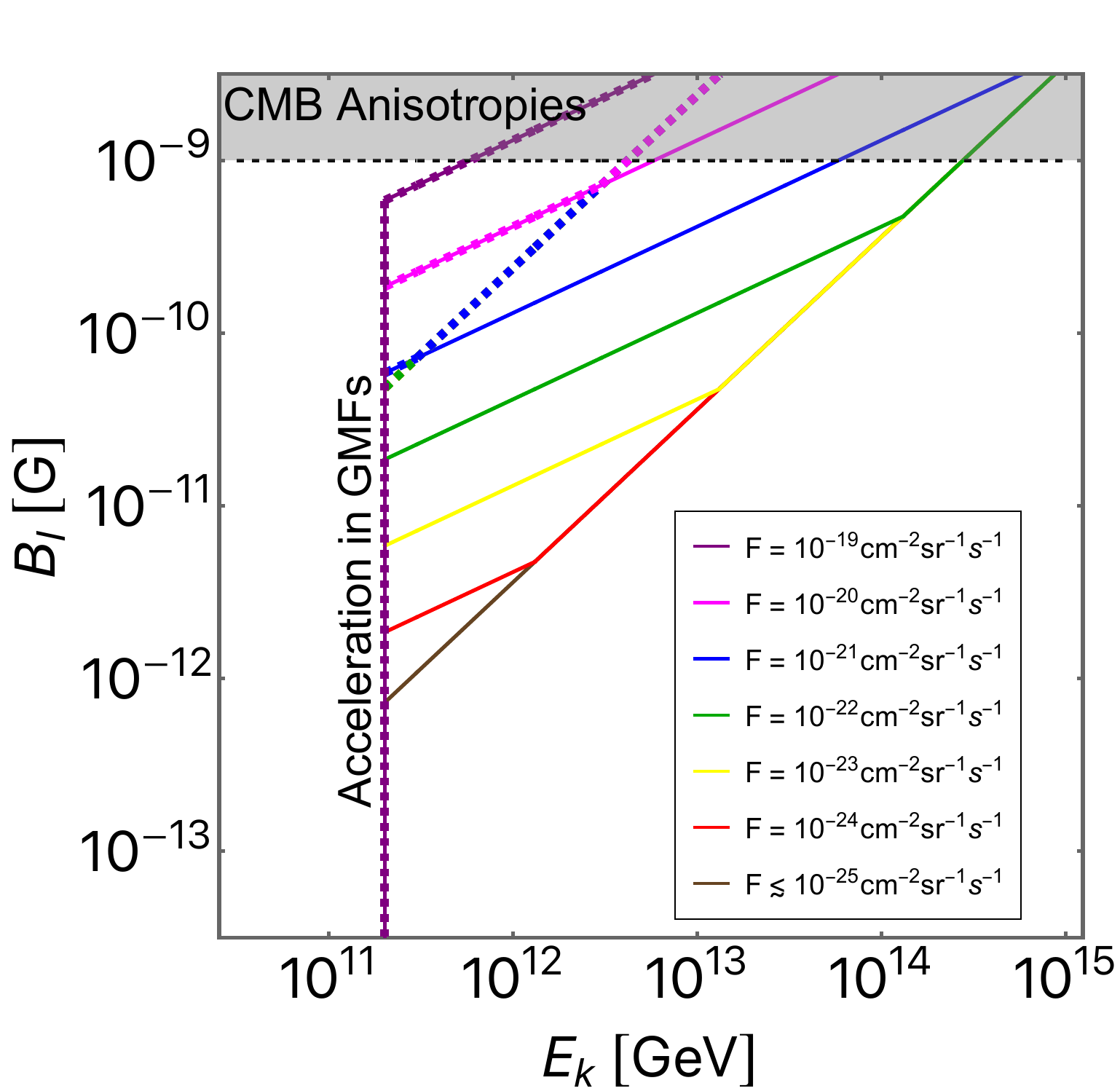}
     \caption{Relation between IGMF strength and kinetic energy of MMs that arrive at Earth with relativistic velocities.
The MM charge is fixed to $g = g_{\mathrm{D}}$, and the line colors denote different values of the MM flux.
Solid lines are for $\lambda_{\rm I} \gtrsim 1/H_0$, while dashed lines are for $\lambda_{\rm I} = 1\, \mathrm{Mpc}$.
The vertical segment where the lines overlap represents the minimum energy acquired in GMFs, and a departure from it signals extra acceleration in IGMFs. One can infer the IGMF strength and coherence length by measuring the MM flux and kinetic energy.
The gray region is excluded by constraints from CMB.}       
\label{fig:test_igmf}
\end{figure}

In \autoref{fig:test_igmf} we focus on MMs that arrive at Earth with relativistic velocities, and show their kinetic energy~$E_{\ro{k}} = m (\gamma_{\ro{E}} - 1)$ in terms of the IGMF strength.
Here the energy is independent of the mass, since
the Lorentz factor of relativistic MMs are set by either one of 
\autoref{eq:2.2}, third line of \autoref{eq:nonhomoGen}, 
second line of \autoref{eq:gamma_v_max}, or \autoref{eq:ek},
which are all inversely proportional to~$m$.
The $E_k$-to-$B_{\ro{I}}$ relation, on the other hand, depends on the MM flux and IGMF coherence length.
Lines with different colors denote different flux values as indicated in the legend, while solid lines are for $\lambda_{\rm I} \gtrsim 1/H_0$, and dashed lines for $\lambda_{\rm I} = 1~\mathrm{Mpc}$. 
(For the dashed lines, flux values of 
$F \lesssim 10^{-22}\, \mathrm{cm}^{-2} \mathrm{s}^{-1} \mathrm{sr}^{-1}$ are collectively described by orange.)
As one moves toward small~$B_{\ro{I}}$, all lines merge into a vertical segment at $E_k \sim 10^{11}\, \ro{GeV}$; this corresponds to the energy acquired in GMFs, and no monopoles are expected to reach the Earth with a smaller kinetic energy.
On the other hand, a larger energy signals extra acceleration in IGMFs,
and for a sufficiently large~$B_{\ro{I}}$
the lines merge to follow
$E_k \sim g B_{\ro{I}} / H_0$ for $\lambda_{\rm I} \gtrsim 1/H_0$,
or $E_k \sim g B_{\ro{I}} (\lambda_{\rm I}/ H_0)^{1/2}$ for $\lambda_{\rm I} < 1/H_0$.
The plot clearly shows that by measuring the flux and kinetic energy of MMs, one not only can infer the existence of IGMFs, but can also measure (a combination of) their strength and coherence length.\footnote{The Telescope Array (TA) \cite{TelescopeArray:2023sbd} recently detected an ultra-high-energy particle with 
$E_{\ro{k}} \approx 2 \times 10^{11}\, \ro{GeV}$.
This may be interpreted as a MM accelerated in the GMF.
Within this MM hypothesis, IGMFs cannot give a larger energy to the MM.
Hence, by crudely assuming the flux to be 
$F \sim 10^{-21}\, \mathrm{cm}^{-2} \mathrm{s}^{-1} \mathrm{sr}^{-1}$
and using \autoref{fig:test_igmf}, 
one can rule out IGMFs of, e.g., 
$B_{\ro{I}} \gtrsim 10^{-10}\, \ro{G}$ with $\lambda_{\ro{I}} \gtrsim 1 \,  \ro{Mpc}$.}

\section{Recasting experimental limits}
\label{sec:results}

\begin{figure}[t!]
    \centering
    \includegraphics[width=\linewidth]{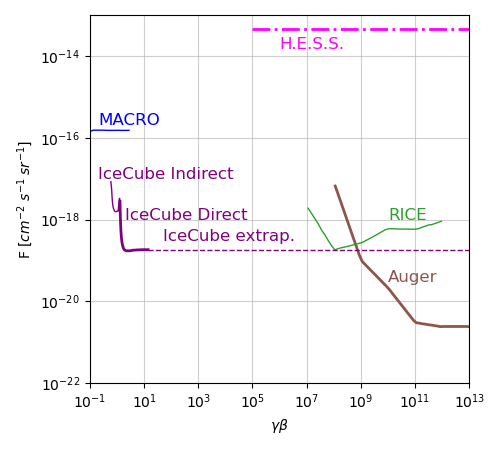}
    \caption{Experimental upper limits on the singly charged ($g=g_{\ro{D}}$) MM flux in terms of the MM velocity, at the detector. 
Shown are limits from MACRO~\cite{MACRO:2002kki} (blue),
IceCube~\cite{IceCube:2021eye} (purple), RICE~\cite{Hogan:2008sx} (green), Auger~\cite{PierreAuger:2016imq} (brown), and a preliminary study~\cite{Spengler:2011} for H.E.S.S. (magenta). 
}
    \label{fig:limits_exp}
\end{figure}

Over the years, numerous experiments have placed constraints on the flux of MMs. See e.g. \cite{Giacomelli:2000de,Balestra:2011ks,Mavromatos:2020gwk} for comprehensive reviews. 
In \autoref{fig:limits_exp}, we collect some of the strongest flux limits,
given in terms of the velocity parameter $\gamma\beta$ at the detector of each experiment,
reported by IceCube~\cite{IceCube:2021eye}, 
RICE~\cite{Hogan:2008sx}, 
MACRO~\cite{MACRO:2002kki}, 
Auger~\cite{PierreAuger:2016imq}, and from a preliminary study for 
H.E.S.S.~\cite{Spengler:2011}. 
One clearly sees that different instruments have sensitivities in different velocity ranges.\footnote{Limits on relativistic MMs that have been obtained by other experiments (e.g. \cite{BAIKAL:2007kno, Detrixhe:2011dan, ANTARES:2025ojl}) are 
comparable to or weaker than the ones we discuss here.
We also note that we do not discuss constraints from searches of nucleon decay catalyzed by MMs, since the catalysis depends on the details of the MM model.} 

\begin{figure}[h!t]
     \centering
     \includegraphics[width=\linewidth]{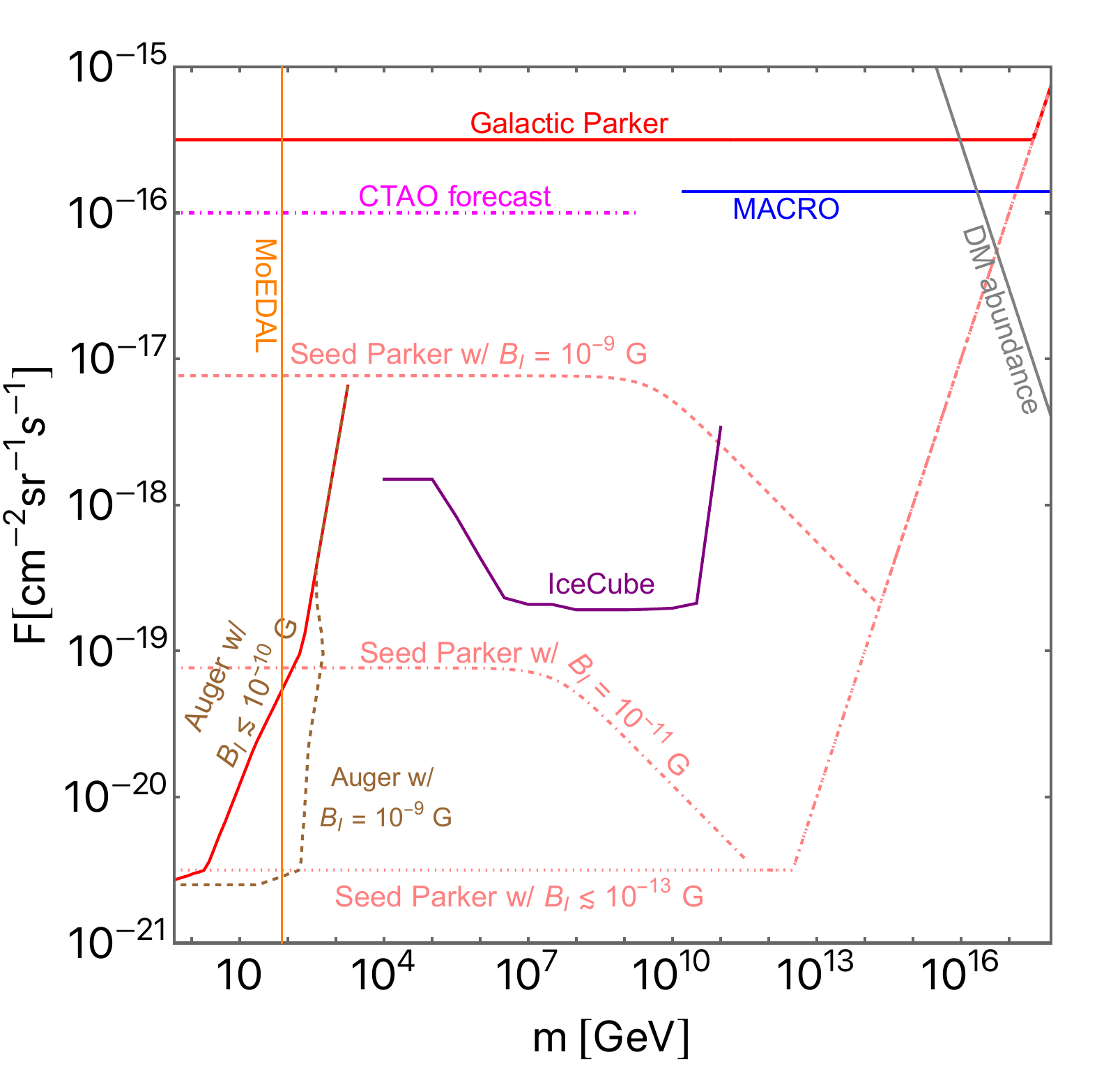}
     \caption{Upper limits on the singly charged ($g=g_{\ro{D}}$) MM flux on Earth, in terms of the MM mass. 
Shown are results from MACRO (blue),
IceCube (purple),
Auger (brown),
expected sensitivity for CTAO (magenta),
Galactic Parker bound (red),
seed Parker bound (pink),
cosmological limit from comparison with the average dark matter density in the universe (gray),
and the mass lower limit from MoEDAL (orange). 
The Auger and seed Parker limits depend on IGMFs, whose field strength is varied. See the text for details.
}\label{fig:results}
\end{figure}

In this section, we recast the experimental limits at the detectors, into limits on Earth as functions of the MM mass. 
For this purpose, we study the energy loss of MMs after their arrival at Earth. Combining this with the acceleration of MMs in cosmic magnetic fields (as described in \autoref{sec:MonAcc}), we evaluate the MM velocity at the detectors as functions of the mass for each experiment. 
The results are reported in \autoref{fig:results}, which represents the main outcome of this work. 

We describe the individual limits in the following subsections. 
Note that for RICE, we only show its limit for reference in \autoref{fig:limits_exp}, but do not perform a recasting since the result is somewhat similar to those from IceCube and Auger.
In the rest of the paper, unless otherwise stated, we consider MMs with a single Dirac charge, i.e. $g = g_{\ro{D}}$. 

\subsection{MACRO} 
\label{sec:macro}

MMs crossing a medium deposit a large amount of energy; the ionization yield of a relativistic unit charge MM is $(g_{\ro{D}}/e)^2 \approx 4700$ times that of a minimum ionizing electrically charged particle~\cite{Ahlen:1978,RevModPhys.52.121, Ahlen:1982}. 
This signature drove the design of the Monopole, Astrophysics and Cosmic Ray Observatory (MACRO), which was a dedicated instrument to search for MMs~\cite{MACRO:2002kki}. It was operated at Gran Sasso INFN National Laboratories (LNGS), at water equivalent depth of 3,800~m, for a decade, until 2000.
The latest global limits based on 9.5 years of data were reported in 
\cite{MACRO:2002jdv},
for MM velocities at the detector within the range of 
$4\cdot10^{-5} \leq \beta \leq 0.99$.
Assuming an isotropic flux at the detector, its upper bound is $F < 1.4\cdot10^{-16}$~cm$^{-2}$s$^{-1}$sr$^{-1}$.
We show this value in \autoref{fig:limits_exp} with a solid blue line.

Velocities of $\beta > 0.99$ (i.e. $\gamma > 7$) were excluded from the MACRO analysis, since such fast-moving MMs could induce showers in the detector that would reduce the efficiency of the analysis.
This implies that the MACRO bound does not apply to light MMs that inevitably obtain large Lorentz factors in cosmic magnetic fields;
however, in the literature it has often been assumed that the bound 
holds for arbitrary MM masses. 

With MACRO's flux sensitivity of order
$10^{-16}$~cm$^{-2}$s$^{-1}$sr$^{-1}$, 
the MM velocity is determined by GMFs,
as we discussed below \autoref{eq:FII}.
Therefore we can use \autoref{eq:ek} to estimate the kinetic energy of the MMs on the Earth's surface, which translates the condition $\beta < 0.99$  into a lower limit on the MM mass: $m \gtrsim 10^{10}\, \mathrm{GeV}$.
This sets the mass cutoff for the MACRO limit.

The MACRO experiment is set underground, 
however we show in \autoref{sec:macro_acceptance} that MMs that have been accelerated in GMFs are energetic enough to reach the detectors from all directions.
Therefore the flux of cosmic MMs at Earth is constrained by 
the MACRO limit at full acceptance,
$F < 1.4\cdot10^{-16}$~cm$^{-2}$s$^{-1}$sr$^{-1}$. 
We show this limit from MACRO in \autoref{fig:results} by the blue line.

\subsection{IceCube} 

Relativistic MMs can emit Cherenkov radiation in a transparent medium such as the Earth's polar ice caps.
The threshold for Cherenkov emission is $\beta > 1/n_r$, where $n_r$ is the refraction index of the medium. In ice, this 
translates into $\beta \gtrsim 0.76$ (i.e. $\gamma\beta \gtrsim 1.2$).
The Cherenkov photon yield of a unit charged MM is $(g_{\ro{D}} n_r/e)^2$ times more than that of a muon with the same velocity~\citep{Tompkins:1965}. 

Located at the South Pole in Antarctica, the IceCube detector is an array of 5160 optical modules equipped with photomultiplier tubes,
arranged in 86 vertical strings deployed into the ice between 1500~m and 2500~m depth, with a total volume of 1~km$^3$. 
Its main scope is the detection of astrophysical and cosmological neutrinos in the ultra-high energy (PeV--UeV) range through the collection of Cherenkov light, emitted by charged leptons that are created by the neutrinos.
Given the signal intensity, a MM event would be very different from a standard neutrino event and therefore relatively simple to identify.

The IceCube analyses~\cite{IceCube:2012,IceCube:2014xnp,IceCube:2015agw,IceCube:2021eye} cover both the direct Cherenkov emission from the MM itself, and indirect Cherenkov emission from secondary ionized electrons.
The latest flux limits at the detector, assuming an isotropic flux, 
are shown in \autoref{fig:limits_exp} by a thin purple curve for 
$0.51 \lesssim \beta \lesssim 0.76$ (from indirect Cherenkov light~\cite{IceCube:2015agw}), and a thick purple curve for $0.76 \lesssim \beta \lesssim 0.995$ (direct Cherenkov light~\cite{IceCube:2021eye}), with $\beta$ being the MM velocity at the detector.
It is claimed in \cite{IceCube:2015agw} 
that the limits can be extrapolated to even larger~$\beta$, for which
the MM signal is expected to be brighter.
The extrapolated limit is shown in \autoref{fig:limits_exp} by a dashed purple line.

The limits at the detector can be translated into those on Earth, by taking into account the MMs' energy loss as they travel through rocks and ice before reaching the detectors.
Such analyses were carried out in IceCube's early works~\cite{IceCube:2012, Christy:2011lza}, but not for the latest limit in~\cite{IceCube:2021eye}. We hence carried out the translation for the latest limit, and further recast them in terms of the MM mass through our acceleration analysis.
The detailed procedure is described in \autoref{sec:icecube_acceptance}, 
and the results are reported in \autoref{fig:results} as the purple line. 
We remark that the shown limit corresponds to the direct Cherenkov and extrapolated limits (i.e. $\beta \gtrsim 0.76$), 
which we find to be applicable to the mass range of
$10^{4}\, \ro{GeV} \lesssim m \lesssim 10^{11}\, \ro{GeV}$.
For the indirect Cherenkov limits, IceCube has not performed an acceptance analysis and hence we did not attempt to do the conversion. 

We remark that IceCube's flux sensitivity saturates the threshold~$10^{-19}\mathrm{cm^{-2}\, sr^{-1}\, s^{-1}}$, below which the acceleration in IGMFs can dominate (cf. discussions below \autoref{eq:FII}).
Hence with further upgrades, IceCube has the potential to probe the IGMF acceleration of MMs.

\subsection{Pierre Auger Observatory} 

Ultrarelativistic MMs deposit large amounts of energy as they traverse the Earth’s atmosphere.
For $\gamma > 10^4$, energy loss in the atmosphere is dominated by photonuclear interactions and pair production, with the deposited energy increasing with $\gamma$ \cite{Wick:2000yc}.
This energy loss results in the generation of secondary electromagnetic showers along the MM trajectory, as well as intense fluorescence emission from atmospheric nitrogen molecules.
Since MMs cross the entire atmosphere, the resulting showers feature much larger energy deposition and deeper development compared to those induced by conventional cosmic rays such as protons.

With an area of $3,000$~km$^2$, Auger is the largest ultra-high-energy cosmic-ray detector currently in operation.
It is provided with a surface-detector array that directly samples the Cherenkov light emission from charged particles in atmospheric showers,
and 24 fluorescence detectors, each of them with a field-of-view of $30^\circ$.
Only the trajectories originating within a $60^{\circ}$ cone around the zenith are selected, 
hence MMs that cross the Earth before emitting the signals are not included in the analysis.

The search for MMs at Auger was reported in \citep{PierreAuger:2016imq}, yielding competitive flux limits in the range $10^{-21}$–$10^{-17}~\mathrm{cm^{-2} sr^{-1} s^{-1}}$, which are shown in Fig.~\ref{fig:limits_exp} as the solid brown curve.
In Fig.~\ref{fig:limits_exp}, the Auger limit is expressed in terms of the Lorentz factor of MMs upon entering the atmosphere,
and is limited to the range $\gamma>10^8$.
In \citep{PierreAuger:2016imq}, Monte Carlo simulations of atmospheric showers were performed for a fixed MM energy of $E_k = 10^{16}~\ro{GeV}$.
However, since energy loss in the ultrarelativistic regime depends on the Lorentz factor and not on~$E_k$, the reported limits are claimed to hold for arbitrary MM energies.

MMs are accelerated in GMFs beyond the Auger threshold of $\gamma = 10^8$ provided that $m \lesssim 10^3~\ro{GeV}$, as shown by the “Auger” dotted lines in \autoref{fig:gammabetam}.
Therefore, the Auger limits apply to this mass range.
However, because Auger is sensitive to fluxes down to
$F \sim 10^{-21}~\mathrm{cm^{-2} sr^{-1} s^{-1}}$,
the precise mass dependence of the constraints can be affected by IGMFs if
$B_{\ro{I}} > 10^{-10}~\ro{G}$ (cf. \autoref{eq:FII}).
In \autoref{fig:results}, we present the Auger limits in terms of MM mass by the brown lines.
The solid brown line corresponds to the case $B_{\ro{I}} \lesssim 10^{-10}~\ro{G}$, where IGMFs are negligible, while the dashed brown line shows the case $B_{\ro{I}} = 10^{-9}~\ro{G}$ with $\lambda_{\ro{I}} \gtrsim 10~\ro{Mpc}$.
The solid and dashed lines converge at $F > 10^{-19}~\mathrm{cm^{-2} sr^{-1} s^{-1}}$.
The dashed line has a non-monotonic shape; 
it should be understood that its left side is excluded.
The origin of this peculiar shape is explained in \autoref{sec:auger_app}.

Finally, we note that IGMFs also impact the constrained MM mass range.
For example, if in the future Auger improves its flux sensitivity uniformly to
$10^{-22}~\mathrm{cm^{-2} sr^{-1} s^{-1}}$,
the corresponding limit would extend up to MM masses as large as
$10^6 \, \ro{GeV}$
in the case of $B_{\ro{I}} = 10^{-9} \ro{G}$ and $\lambda_{\ro{I}} \gtrsim 1/H_0$ (see \autoref{fig:vEarth9}).

\subsection{Forecast for CTAO}

Imaging Atmospheric Cherenkov Telescopes (IACTs) are large parabolas designed to detect Cherenkov light produced by primary gamma rays and cosmic rays in the Earth’s atmosphere.
For MMs, the Cherenkov yield would be extremely bright (cf. the subsection on IceCube) and visible along the entire MM trajectory.
Near the ground, the refractive index of air is $n_{\rm r} \sim 1+10^{-4}$, which sets the Cherenkov threshold at $\gamma \gtrsim 70$ \cite{Spengler:2009}.
At higher altitudes, such as 30~km, the reduced atmospheric density lowers the refractive index to $n_{\rm r} \sim 1+10^{-6}$ \cite{USStandardAtmosphere:1976dan}, raising the threshold to $\gamma \gtrsim 700$.
Thus, faster MMs would begin emitting Cherenkov light at higher altitudes.

In the atmosphere, the Cherenkov emission angle, given by $\cos\theta = 1/(n_{\rm r} \, \beta)$, increases with atmospheric depth for a fixed $\beta$.
This effect produces a distinctive imaging signature in IACTs. For fast MMs passing close to the telescopes, a double image would appear --- one from Cherenkov photons emitted at small angles in the upper atmosphere, and another from photons emitted much closer to the ground.
At intermediate depths, the Cherenkov light would focus outside the telescope camera.
Such a characteristic image would make MMs difficult to confuse with ordinary cosmic rays.

The possibility of searching for MM-induced Cherenkov light with IACTs has been studied for the High Energy Stereoscopic System (H.E.S.S.) experiment, which is an array of IACTs.
A forecast was presented in \cite{Spengler:2011}, while five years of H.E.S.S. data was analyzed in \cite{Spengler:2009}, resulting in a flux limit of $F \lesssim 4.5 \times 10^{-14}~\mathrm{cm^{-2} s^{-1} sr^{-1}}$, shown as the dot-dashed magenta line in \autoref{fig:limits_exp}.
This limit applies to MMs entering the atmosphere with $\gamma \gtrsim 10^5$, since \cite{Spengler:2009} only considered cases where Cherenkov light is emitted at sufficiently high altitudes.
Extending the analysis to smaller $\gamma$ values requires the generation of computationally expensive Monte Carlo simulations to evaluate the effective detector area, but such computations were not feasible 
due to limited computing resources for the study in~\citep{Spengler:2009}.
At very high $\gamma$, atmospheric energy loss of MMs becomes significant, though this effect can be ignored since it only affects MM masses already excluded by MoEDAL (cf. \autoref{fn:1}).

The main limitation of IACTs is their relatively narrow fields of view, which greatly reduce acceptance compared to observatories like Auger or IceCube.
Nonetheless, given that the H.E.S.S. analysis \cite{Spengler:2009} was based on a simplified approach, and in light of the upcoming next-generation experiment Cherenkov Telescope Array Observatory (CTAO)~\cite{CTAConsortium:2017dvg}, it is important to consider the future prospects of IACTs for MM searches.

Unlike H.E.S.S., which consists of a single array of five telescopes, CTAO will include two arrays — one in the northern hemisphere with 13 telescopes, and one in the southern hemisphere with 70 telescopes.
Consequently, CTAO will have a field of view of $10^\circ$, which, compared to $6^\circ$ for H.E.S.S., corresponds to larger coverage by a factor of $(10^\circ / 6^\circ)^2 \approx 3$.
Moreover, its effective area is expected to be larger by a factor of $\sim 10$~\cite{Gammapy:2023gvb}.
Furthermore, CTAO is planned to operate for 30 years, i.e., six times longer than the H.E.S.S. dataset analyzed in \cite{Spengler:2009}.
Altogether, this implies a net improvement in flux sensitivity by a factor of roughly
$3 \times 10 \times 6 \approx 200$,
leading to an estimated sensitivity of
$\sim 10^{-16}~\mathrm{cm^{-2} s^{-1} sr^{-1}}$.
This forecast for CTAO is shown in \autoref{fig:results} as the dot-dashed magenta line.

For CTAO, we adopt a velocity threshold of $\gamma \gtrsim 70$, under the assumption that accurate modeling of the detector’s effective area will allow sensitivity to Cherenkov photons emitted close to the ground.
With the estimated flux sensitivity, CTAO can only probe MM acceleration in GMFs.
Thus, using Eq.~\eqref{eq:ek}, the velocity threshold (shown in \autoref{fig:gammabetam} by the black dotted lines labeled “IACTs”) translates into $m \lesssim 10^{9}~\mathrm{GeV}$, which sets the cutoff of the CTAO limit in \autoref{fig:results}.

Although CTAO will generally be less sensitive than Auger or IceCube, it will provide an independent and complementary search channel for MMs.
In particular, CTAO has the potential to surpass the Galactic Parker bound, and explore the mass gap of $10^{3}~\mathrm{GeV} \lesssim m \lesssim 10^{4}~\mathrm{GeV}$ left between the Auger and IceCube constraints.

\subsection{MoEDAL}
Heavy-ion collisions at the LHC produce extremely strong magnetic fields. 
MMs produced by the magnetic dual of the Schwinger effect~\cite{Schwinger:1951nm,Affleck:1981ag,Affleck:1981bma,Gould:2017zwi,Gould:2019myj}\footnote{Although $g_{\ro{D}} \gg 1$, the Schwinger production rate 
can be calculated using an instanton method~\cite{Affleck:1981ag,Affleck:1981bma}.
On the other hand for Drell--Yan processes~\cite{CMS:2021rwb, ATLAS:2023esy, MoEDAL:2021mpi}, there are considerable theoretical uncertainties in the MM production cross section.}
are searched for by the Monopole and Exotics Detector at the LHC
(MoEDAL)~\cite{MoEDAL:2009jwa,MoEDAL:2021vix}.\footnote{Recently a similar search was also carried out with ATLAS~\cite{ATLAS:2024nzp}. Schwinger limits on the MM mass can also be obtained from magnetars \cite{Gould:2017zwi,Hook:2017vyc} and primordial magnetic fields \cite{Kobayashi:2021des,Kobayashi:2022qpl}.}
MoEDAL is a largely passive detector, comprised of nuclear track detectors and magnetic trappers for tracking and capturing the produced MMs.
The results so far are compatible with no MM production, 
based on which a lower limit on the mass $m > 75\, \mathrm{GeV}$ 
has been obtained
for singly charged ($g=g_{\ro{D}}$) MMs~\cite{MoEDAL:2021vix}.
The lower mass limit from MoEDAL is represented by the orange vertical line in \autoref{fig:results}. This limit of course is unaffected by cosmic magnetic fields, however we show it for completeness.

\subsection{Parker bounds}

Among the indirect MM constraints, the most notable is the so-called Parker bound, which is obtained from the requirement that the MMs should not short out the GMFs~\cite{Parker:1970xv,Turner:1982ag}.
For MMs that propagate to Earth from outside the Milky Way,
the Parker bound on their flux takes the form~\cite{Kobayashi:2023ryr}\footnote{Even if the MMs' velocity at Earth is well described by its mean value (cf. \autoref{fn:2}), the variance of the MMs' energy loss inside the Galaxy may be nonnegligible. This, along with other discussions require a few conditions for the Parker bound in \autoref{eq:parker} to hold; see Eqs.~(2.16)--(2.18) in \cite{Kobayashi:2023ryr}. These conditions are satisfied for the parameters shown in \autoref{fig:limits_exp}.}
\begin{equation}
\label{eq:parker}
\begin{aligned}
 F \lesssim \ro{max.} \Biggl\{
&\frac{10^{-16}}{\mathrm{cm}^{2} \, \mathrm{s} \, \mathrm{sr}}
 \left(\frac{m \left(\gamma_{\ro{MW}} - 1 \right)}{10^{11}\, \mathrm{GeV}} \right) \left( \frac{g}{g_{\mathrm{D}}} \right)^{-2} 
\left( \frac{\lambda_{\mathrm{G}}}{1 \, \mathrm{kpc}} \right)^{-1}
\left(\frac{\tau_{\mathrm{gen}}}{10^{8}\, \mathrm{yr}} \right)^{-1},
\\
 &\frac{10^{-16}}{\mathrm{cm}^{2} \, \mathrm{s} \, \mathrm{sr}}
\left( \frac{g}{g_{\mathrm{D}}} \right)^{-1}
\left( \frac{B_{\mathrm{G}}}{10^{-6}\, \mathrm{G}} \right)
\left( \frac{R}{\lambda_{\mathrm{G}}} \right)^{1/2}
\left(\frac{\tau_{\mathrm{gen}}}{10^{8}\, \mathrm{yr}} \right)^{-1}
\Biggr\} ,
\end{aligned}
\end{equation}
for a generic magnetic charge~$g$.
Here, $\gamma_{\ro{MW}} = (1- v_{\ro{MW}}^2)^{-1/2}$ is the Lorentz factor of the MMs with respect to the Milky Way,\footnote{The Parker bound
also applies to MMs that are clustered with the Milky Way (this can happen 
for singly charged MMs if $m \gtrsim 10^{18}\, \ro{GeV}$), for which $v_{\ro{MW}}$ becomes the virial velocity $\sim 10^{-3}$. We also note that since the velocities of both clustered and unclustered MMs with respect to the Milky Way are comparable to or larger than that of the Earth, the MM flux in the rest frames of the Milky Way and Earth are the same order.} which is given as 
$v_{\ro{MW}} = \max\{v_{\mathrm{I}}, v_{\rm p}\}$
in terms of the IGMF-induced velocity in \autoref{eq:v_MW} 
and the Milk Way's peculiar velocity.
The other parameters are for the GMF, with 
$\tau_{\mathrm{gen}}$ being the dynamo time scale 
which is often considered to be $\sim 10^8\, \ro{yr}$.

The original works~\cite{Parker:1970xv,Turner:1982ag} derived the bound from the survival of the present-day GMFs, whose amplitude is $B_{\rm G} \sim 10^{-6} \, \mathrm{G}$. 
This discussion can also be applied to the initial seed magnetic fields of the Milky Way~\cite{Adams:1993fj}.\footnote{Parker-type bounds can also be derived from fields in galactic clusters \cite{Rephaeli:1982nv}, and in the primordial universe \cite{Long:2015cza, Kobayashi:2022qpl};
see \cite{Kobayashi:2023ryr} for a comprehensive discussion.} The resulting bound takes the same form as above, but becomes stronger since the seed fields should have been much weaker than the present-day fields; however it should also be noted that the inferred seed field strength has a large uncertainty, ranging typically between
$10^{-30}\, \ro{G} \lesssim B_{\ro{G}} \lesssim 10^{-10}\, \ro{G}$~\cite{Widrow:2002ud}.
In \autoref{fig:results}, the red line shows the original Parker bound based on the present-day GMFs, while the pink lines show the bounds from seed fields whose amplitude is assumed here as
$B_{\ro{G}} = 10^{-11}\, \ro{G}$.

For the Parker bound, the entire Milky Way serves as the detector, so unsurprisingly the constraint depends on the MMs' incident velocity on the Milky Way.
The bound becomes weaker for a larger Lorentz factor, because
fast-moving MMs pass through the Galaxy while being minimally deflected by the GMFs, hence being less effective in 
dissipating the magnetic energy. 
Hence in order to correctly estimate the Parker bound, one needs to take into account the acceleration of MMs in IGMFs.
It was shown in \cite{Perri:2023ncd} that, 
although the Parker bound from the present-day GMFs is barely affected by observationally viable IGMFs,
the bound from seed fields depends sensitively on the IGMF strength.
In \autoref{fig:results}, the seed Parker bound is shown for
$B_{\rm I}= 10^{-9}\, \mathrm{G}$ (pink dashed), $B_{\rm I} = 10^{-11} \, \mathrm{G}$ (dotdashed), and $B_{\rm I} \lesssim 10^{-13}\, \mathrm{G}$ (dotted).\footnote{Note that the seed Parker bound depends on the IGMF even if its flux limit does not reach the value in \autoref{eq:FII}.}
The bounds here are independent of the IGMF correlation
length, as long as it is as large as 
$\lambda_{\ro{I}} \gtrsim 10^{-5}\, \ro{Mpc}$.

\subsection{Cosmic abundance bound}

For MMs that are nonrelativistic in the intergalactic space, i.e. $v_{\ro{I}} \ll 1$, their abundance should not exceed the average dark matter density in the universe:
$\rho_{\ro{DM}} \approx 1.3 \times 10^{-6} \, \ro{GeV} \, \ro{cm}^{-3}$~\cite{Planck:2018vyg}.
The MM density is written as $\rho = m n$
in terms of the number density~$n$, which is also related to the intergalactic flux as
$F_{\ro{I}} = n v_{\ro{I}} / 4 \pi $.
Considering MMs that pass through the Milky Way without being captured
(i.e. $m \ll 10^{18}\, \ro{GeV}$), 
their flux on Earth is the same as the incident flux on the Milky Way, which is further related to the intergalactic flux in the CMB rest frame via
$F = F_{\ro{I}} \, \ro{max.} \{ 1, v_{\ro{p}} / v_{\ro{I}} \} $.
Hence the abundance limit constrains the flux at Earth as,
\begin{equation}\label{eq:cosmological}
\begin{split}
 F &< \frac{\rho_{\ro{DM}} }{4 \pi m} 
\, \ro{max.} \{ v_{\ro{I}}, v_{\ro{p}}  \} \\
&\approx \frac{3 \times 10^{-17}}{\mathrm{cm}^{2}\,\mathrm{s}\,\mathrm{sr}}  \left(\frac{m}{10^{17}\, \ro{GeV}}\right)^{-1} 
\left[ \frac{ \ro{max.} \{ v_{\ro{I}}, v_{\ro{p}} \} }{10^{-3}} \right].
\end{split}
\end{equation}
Notice that the right-hand side also depends on the flux through 
$v_{\ro{max}}$ (cf. \autoref{eq:gamma_v_max}). The flux variable here can be identified with $F$ on Earth, from a similar argument given below \autoref{eq:vEarth}. 

The abundance bound is shown in \autoref{fig:results} as the gray line, which is displayed for masses of
$10^{15}\, \ro{GeV} \lesssim m < 10^{18}\, \ro{GeV}$.
Within this mass range, the abundance bound is independent of the IGMF parameters
as long as $B_\ro{I} \lesssim 10^{-9}\, \ro{G}$, 
since $v_\ro{I} < v_{\ro{p}} \sim 10^{-3}$ holds for flux values that saturate the bound.
We stress that the bound of \autoref{eq:cosmological}
applies to MMs that are nonrelativistic and are not clustered in the Galaxy, which is the case within the displayed mass range. 
Lighter MMs that obtain relativistic velocities in the IGMF contribute instead to the extra radiation density of the universe.

\section{Discussion and Conclusions}
\label{sec:concl}

In this work, we reported experimental limits on the flux of MMs 
in terms of the mass, by 
self-consistently estimating the MMs' acceleration in cosmic magnetic fields, and the deceleration in Earth, until they reach the detectors.
Our methodology for translating velocity-dependent limits into mass-dependent limits can be easily applied to any other experiment.

The limits are summarized in \autoref{fig:results}.
For MMs with $m \gtrsim 10^{11}\, \ro{GeV}$, 
the only terrestrial experiment setting a constraint is MACRO. In this mass range, astrophysical limits such as the seed Parker bound and the cosmic abundance bound, are stronger than the MACRO limit by orders of magnitude.
For $m \lesssim 10^{11}\, \ro{GeV}$, 
Auger and IceCube currently provide the strongest limits among the terrestrial experiments. 
In this mass range, the seed Parker bound strongly depends on the IGMF strength; in particular with $B_{\ro{I}} \gtrsim 10^{-11}\, \ro{G}$, it becomes comparable to or weaker than the Auger and IceCube limits.
We also note that IACTs can fill the mass gap between Auger and IceCube.
MMs with $m \lesssim 10^{2}\, \ro{GeV}$ are excluded by MoEDAL.

Our results explicitly show that experiments targeting high-energy cosmic rays are very powerful in studying intermediate to low mass MMs.
This is because both GMFs and IGMFs give large kinetic energies to cosmic MMs.
As many cosmic-ray experiments are under construction, or will soon release more data, we expect the flux bounds to further improve in the near future.

We also showed that limits from terrestrial experiments can be affected by IGMFs, if the flux sensitivity reaches below
$ 10^{-19}\, \mathrm{cm}^{-2} \mathrm{s}^{-1} \mathrm{sr}^{-1} $.
This in turn opens up the possibility of probing 
IGMFs through their effects on the MM acceleration.
In particular, a measurement of the flux and kinetic energy of MMs on Earth could provide information about the strength and coherence length of the IGMFs.
Auger already has the necessary flux sensitivity for probing the IGMF acceleration, while IceCube is close to achieving it.
Our results stress the importance of IGMF studies for probing MMs, and vice versa. 

Our analysis of the MM acceleration in cosmic magnetic fields can further be improved, for instance, by using more accurate GMF models such as those of \cite{Unger:2023lob},
and numerically computing the MM dynamics therein. 
We also remark that there can be additional magnetic fields 
in the extragalactic space, such as those in cosmic filaments, and fields transported by galactic winds,
which we did not take into account. 
We leave these studies for the future.

\paragraph{Acknowledgement}
We thank Kyrilo Bondarenko for initial collaboration.
We are also grateful to Ivan De Mitri, Fausto Guarino, Mariia Khelashvili, Andrew Long, Pranjal Ralegankar, Qaisar Shafi, and Piero Ullio for helpful discussions. We thank the anonymous journal referee for useful comments.
The work of T.K. was supported in part by the European Union - NextGenerationEU through the PRIN Project ``Charting unexplored avenues in Dark Matter'' (20224JR28W). 
T.K. also acknowledges support from INFN TAsP and JSPS KAKENHI (JP22K03595). 
MD acknowledges support from the Italian MUR Departments of Excellence grant 2023-2027 “Quantum Frontiers”.
D.P. was partially supported by the National Science Centre, Poland, under research grant no. 2020/38/E/ST2/00243.

\appendix

\section{Detector acceptance of MACRO}
\label{sec:macro_acceptance}

\begin{figure}[t!]
    \centering    
    \includegraphics[width=\linewidth]{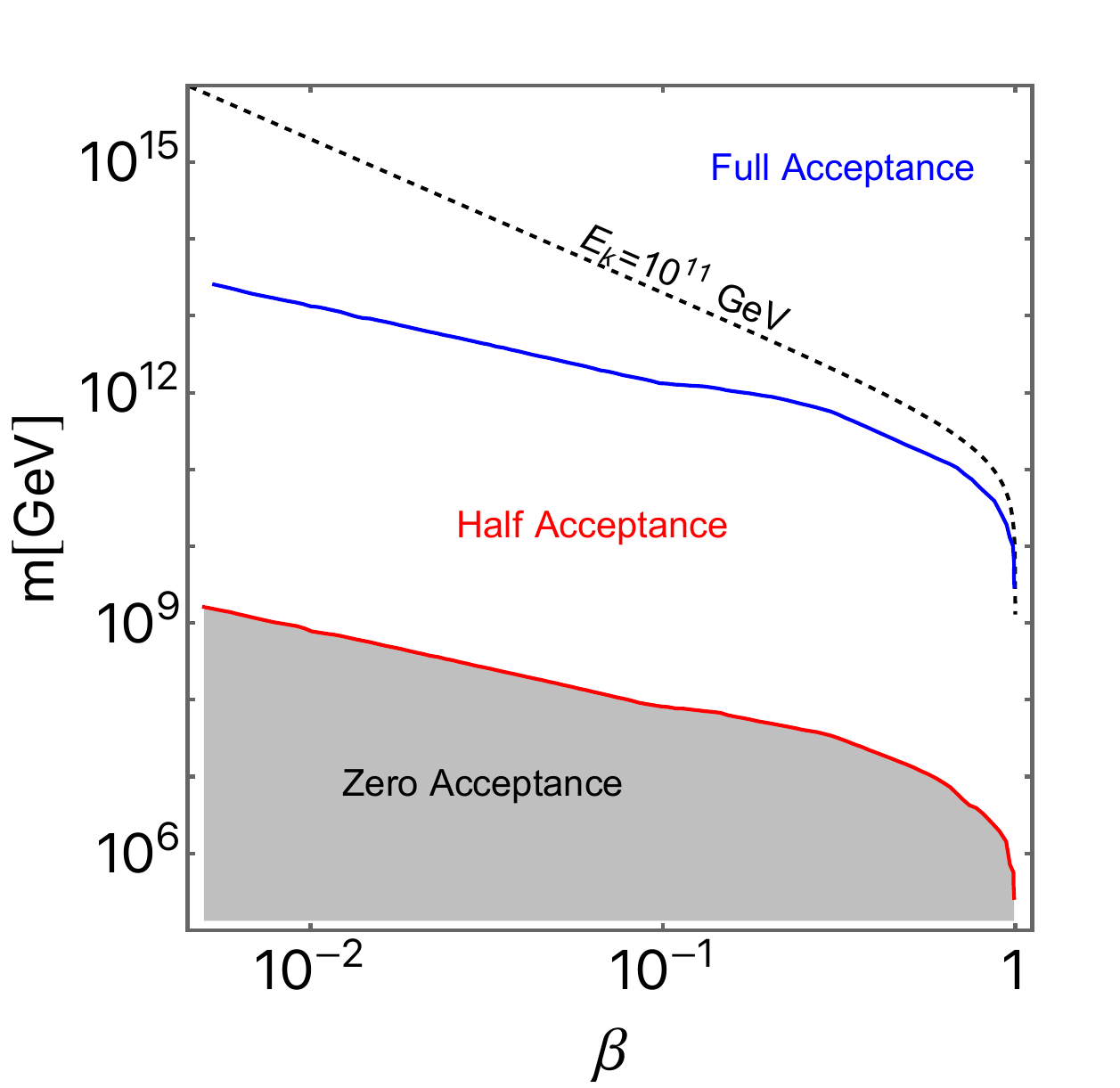}
    \caption{The acceptance of MACRO in the plane of the MM mass and 
velocity at the Earth surface. 
MMs reach the detector from all directions in the parameter region above the blue line, MMs reach only from above in the region between the blue and red lines, and no MMs reach the detector in the gray shaded region.
The dashed black line shows the velocity-mass relation for MMs accelerated in GMFs. This figure is based on Fig.~6.18 of \cite{demitri}.}
    \label{fig:macro_gmf}
\end{figure}

As the MACRO detectors are located beneath the Gran Sasso mountain, their acceptance depends on the incoming direction of the MMs, which cross different portions of the Earth and thus lose different amounts of energy before reaching the detector.\footnote{The energy loss in Earth's atmosphere is smaller than $10^{11}\, \ro{GeV}$ for MMs with both $\gamma \lesssim 10^{10}$ and $m \gtrsim 100\, \ro{GeV}$~\cite{Wick:2000yc}. 
At larger $\gamma$, photonuclear interactions induce larger losses;
at smaller $m$, bremsstrahlung losses can also be nonnegligible.
For MMs accelerated in cosmic magnetic fields, 
by writing $\gamma = \gamma_{\ro{E}}$ (cf. \autoref{eq:vEarth}) as a  function of~$m$, one can check that with observationally consistent IGMFs, 
photonuclear and bremsstrahlung losses in air are negligible compared to the energy gain from GMFs and IGMFs, as long as
$m \gtrsim 100\, \ro{GeV}$.
\label{fn:1}}
The acceptance of MACRO is reported in \cite{demitri,Derkaoui:1998uv}. 
Based on their results,\footnote{The mass and velocity thresholds in \autoref{fig:macro_gmf} are taken from 
Fig.~6.18 in \cite{demitri} and Tabs. 1 to 4 in \cite{Derkaoui:1998uv}.
Note, though, that for some unknown reason, the thresholds shown in Tab.~2 of \cite{MACRO:2002jdv} do not agree with the values in \cite{demitri,Derkaoui:1998uv}, even though these works are cited in \cite{MACRO:2002jdv}.}
we show in \autoref{fig:macro_gmf} 
the three distinctive cases in the plane of the MM mass and velocity at the Earth surface:
In the region above the blue line, 
the MMs have large enough kinetic energy to cross the Earth, hence their flux at the detector is isotropic and is limited as
$F < 1.4 \times 10^{-16}$~cm$^{-2}$s$^{-1}$sr$^{-1}$
for velocities up to $\beta \leq 0.99$ (see \autoref{sec:macro}).
In the region between the blue and red lines, 
the MMs only have energy to cross the mountain above MACRO but not the entire Earth, hence the flux limit on Earth is weaker by a factor of two.
In the gray shaded region, the MMs cannot even cross the mountain and thus are unconstrained. 

In the figure, we also show the  velocity-mass relation for MMs accelerated by GMFs (cf. \autoref{eq:ek}) 
by the black dashed curve.
One sees that just with GMFs, the MMs obtain enough energy to cross the Earth and reach the MACRO detectors. 
We thus find that the full detector acceptance can be exploited, 
and MMs arriving on Earth are generically constrained as
$F < 1.4\times10^{-16}$~cm$^{-2}$s$^{-1}$sr$^{-1}$.

\section{Detector acceptance of IceCube}
\label{sec:icecube_acceptance}

In this appendix, we discuss the acceptance of the IceCube detectors, which we subsequently use to recast the latest IceCube limits in terms of the MM mass.

\subsection{Detection efficiency}

In order to be detected by IceCube, a MM needs to have a large enough kinetic energy at the Earth's surface so that it can reach the detector with a relativistic velocity. The acceptance of IceCube also depends on the incoming direction, since MMs lose a maximum energy by crossing the full length of the Earth, while the energy loss is smaller from slanted directions.
Moreover, upgoing MMs that have passed through the Earth are easier to discriminate from the atmospheric muon background than downgoing ones, since the muons can only reach the detector from above.
In order to take into account all such effects, 
we introduce the detection efficiency~$\epsilon$ ($\leq 1$) as the ratio between the flux of MMs that can be detected by the IceCube detector, and the flux at the Earth's surface.
Then the flux limit on Earth is given by 
$1/\epsilon$ times the limit at the detector obtained by assuming an isotropic flux. 

\begin{figure}[t]
    \centering    
    \includegraphics[width=0.9\linewidth]{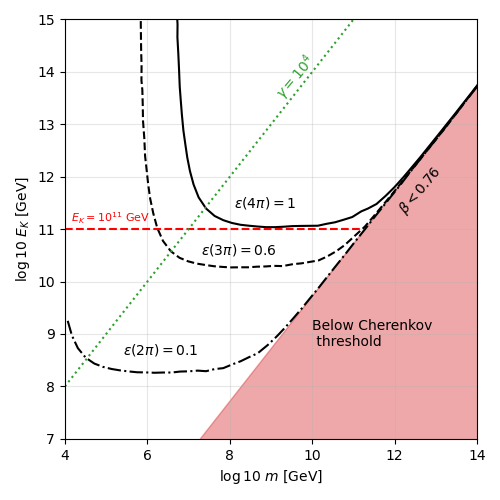}
    \caption{Detection efficiency of IceCube in terms of the MM mass and kinetic energy at the Earth's surface. 
The MMs that are actually constrained by IceCube are those along the 
$E_k \sim 10^{11}\, \ro{GeV}$ line.
This figure is based on Tab.~F.2 of \cite{Christy:2011lza} and Fig.~11 of \cite{IceCube:2012}. See the text for details.}
    \label{fig:ic_boost1}
\end{figure}

The acceptance was studied in detail in IceCube's early works~\cite{Christy:2011lza,IceCube:2012}
which analyzed about 1 year of data.
In particular, the acceptance solid angle is given in Fig.~11 of \cite{IceCube:2012}, and the efficiency~$\epsilon$ can be read off from Tab.~F.2 of \cite{Christy:2011lza}.\footnote{We obtained $\epsilon$ from Tab.~F.2 of \cite{Christy:2011lza} as the ratio between the 
``Final Limit'' and its value at full acceptance, 
$F(4\pi) = 3.382 \times 10^{-18} \, \mathrm{cm^{-2}sr^{-1}s^{-1}}$. 
We then associated $\epsilon$ with the acceptance angle from the
``$\cos \theta$ bins''. The values have mild dependence on $\gamma$, which we neglected.}
We show the values of the acceptance angle and $\epsilon$ in \autoref{fig:ic_boost1}, in the plane of the MM mass and kinetic energy at the Earth's surface.
In the region above the solid black curve, MMs can be detected from any direction (i.e. $4\pi$~sr acceptance) and the efficiency is $\epsilon = 1$.
On the other hand, as one goes below the solid curve, the efficiency decreases. Along the dashed (dot-dashed) black curve, the acceptance angle is $3\pi$ ($2 \pi$)~sr, and the 
efficiency is $\epsilon \approx 0.6$ ($0.1$). 
We stress that $\epsilon$ depends also on how well the signal can be discriminated from the background, and hence it does not scale linearly with the acceptance angle.

The shape of the $\epsilon$~contours can be understood as follows.
Firstly, in the shaded region in the plot, the velocity at the Earth's surface is already smaller than the Cherenkov threshold in ice ($\beta = 0.76$); hence even if such MMs reach the detector without losing any energy, they do not directly emit Cherenkov light.
Secondly, MMs with $\gamma < 10^4$ (i.e. below the green dotted line)
lose energy mainly through atomic excitations and ionization losses. These are collectively referred to as collisional energy losses and are largely insensitive to the MM mass~\cite{Ahlen:1978}, thus 
yielding the plateau region of the $\epsilon$~contours.
In particular, the collisional loss through Earth for relativistic MMs\footnote{Note that in \autoref{sec:macro_acceptance} the energy loss of nonrelativistic MMs through Earth is discussed. See also \cite{Derkaoui:1998uv} for detailed discussions.} 
is $\Delta E_k \sim 10^{11}\, \ro{GeV}$, which sets the low-energy cutoff of the $4\pi$-acceptance region. 
Thirdly, MMs with $\gamma > 10^4$ (i.e. above the green dotted) lose energy mainly  through pair production and photonuclear interactions. 
The energy loss through these processes scale almost linearly with the initial energy, i.e. 
$\Delta E_k 
\propto E_k $~\cite{Wick:2000yc,Detrixhe:2011dan}.
Consequently, the $\epsilon$~contours exhibit a low-mass cutoff.
We also remark that the MMs' energy loss in the atmosphere is negligible for IceCube, cf. \autoref{fn:1}.

\subsection{Conversion of latest limits}

The latest IceCube analysis~\cite{IceCube:2021eye} using 8~years of data
only report limits at the detector,\footnote{Specifically, only mass lower limits for MMs with $\beta \sim 0.76$
to reach the detector is presented in \cite{IceCube:2021eye}.} 
which are shown in \autoref{fig:limits_exp} as the ``IceCube Direct'' and ``extrap.'' lines.
Let us translate them into limits at the Earth's surface using the detection efficiency we read off from IceCube's earlier analyses
(assuming that the efficiency did not change significantly during the years). By further considering the MM velocity at the Earth's surface to have been sourced in cosmic magnetic fields, we then recast the limits in terms of the MM mass. 

With the flux values constrained by IceCube being of $10^{-19}$~cm$^{-2}$s$^{-1}$sr$^{-1}$ or larger, 
the MMs are predominantly accelerated in GMFs and acquire an energy of $E_k \sim 10^{11}$ GeV (cf. \autoref{eq:ek}).
This value is shown in \autoref{fig:ic_boost1} by the red dashed horizontal line. 
We are thus constrained to be on this line, along which $\epsilon$ is a unique function of $m$, or alternatively, of $\beta$ at the Earth's surface. 
Moreover, we consider the velocity of MMs reaching the detector to have minimally changed from its value at the Earth's surface. 
Then we can evaluate the flux limit on Earth by 
dividing the limit at the detector by the efficiency
as $F_{\ro{Earth}} (\beta) = F_{\ro{detector}} (\beta) / \epsilon (\beta)$, 
and further converting $\beta$ into $m$ via \autoref{eq:ek}. 

\begin{figure}[t]
    \includegraphics[width=0.9\linewidth]{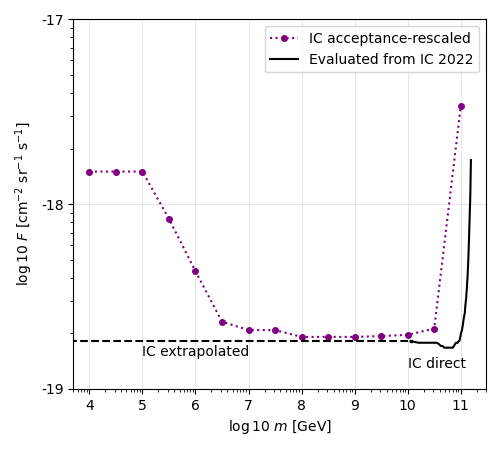}
    \caption{The purple dotted curve shows the latest IceCube limit~\cite{IceCube:2021eye} on the MM flux at Earth, as a function of mass. The black curve shows the limit derived without taking into account the detection efficiency.}
    \label{fig:ic_boost2}
\end{figure}

The final limits are shown in \autoref{fig:ic_boost2} by the purple dashed line. 
We carried out the computation for a discrete set of $m$~values with an interval of 0.5 dex, as the efficiency is provided in this way in \cite{Christy:2011lza}.
The black line shows an intermediate step, where the 
direct Cherenkov and extrapolated limits at the detector are rewritten in terms of $m$ via \autoref{eq:ek}, but not divided by~$\epsilon$. 
The limits have a high-mass cutoff at $m \sim 10^{11}\, \ro{GeV}$, which corresponds to the 
Cherenkov threshold $\beta = 0.76$, as one can read off from
\autoref{fig:ic_boost1} at the intersection between the red dashed line and the shaded region. 
For masses within $10^8-10^{10}$~GeV, 
the $\epsilon$~parameter is almost unity and thus the final limits are only slightly affected by the detection efficiency.
As one goes to even smaller masses the limit becomes weaker due to the decreasing efficiency, and the limit effectively disappears 
at $ m \lesssim 10^4\, \ro{GeV}$ where the efficiency reduces to $\epsilon \lesssim 0.1$. 
The limit is seen to become flat at $m \leq 10^5\, \ro{GeV}$ because, for some unknown reason, the efficiency in \cite{Christy:2011lza} does not reveal any $m$~dependence there.

The IceCube collaboration has not provided studies of the detector acceptance for secondary Cherenkov emission from MMs with $0.51 \lesssim \beta \lesssim 0.76$ (cf. thin solid purple line in \autoref{fig:limits_exp}).
Hence we do not attempt to recast limits in this range.

\section{Effects of intergalactic magnetic fields on Auger limits}
\label{sec:auger_app}

\begin{figure}[t!]
    \centering    
    \includegraphics[width=\linewidth]{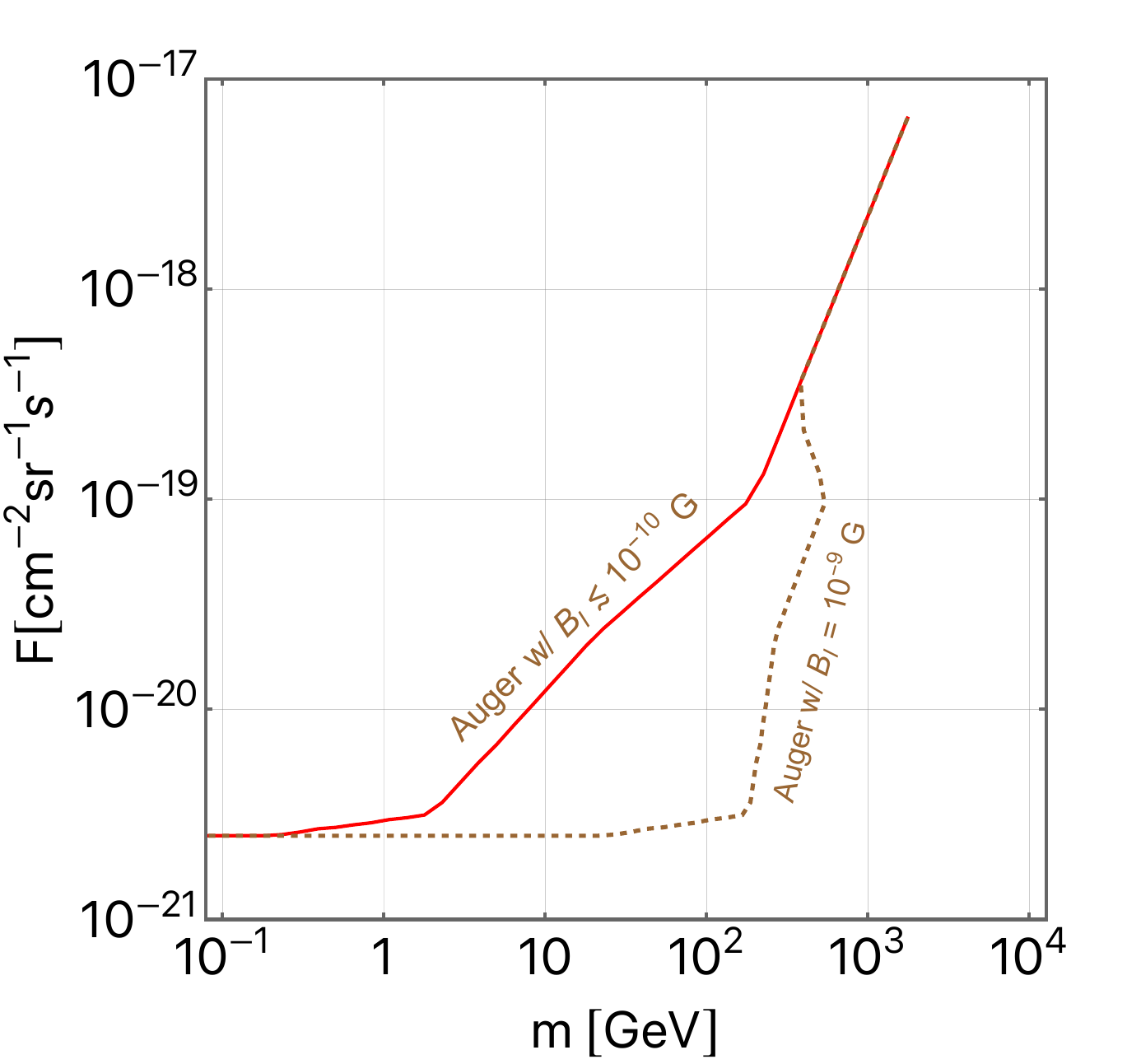}
    \caption{Zoom-in of the Auger limits in \autoref{fig:results}. 
The solid line is for $B_{\ro{I}} \lesssim 10^{-10}\, \ro{G}$, 
while the dashed line is for 
$B_{\ro{I}} = 10^{-9}\, \ro{G}$ and $\lambda_{\ro{I}} \gtrsim 1/H_0$.}
    \label{fig:auger_igmf}
\end{figure}

The flux limits by Auger can have a non-monotonic mass dependence in the presence of IGMFs, as shown in \autoref{fig:results}, and also magnified in \autoref{fig:auger_igmf}. 
We explain this behavior in this appendix.

The Auger limit as a function of the Lorentz factor is shown in \autoref{fig:limits_exp}, and it approximately takes the form of a doubly broken power law. 
If the MMs are predominantly accelerated in GMFs to ultrarelativistic velocities, the Lorentz factor is related to the mass via
$\gamma \propto 1/m$, cf. \autoref{eq:ek}.
The limit is hence mirrored when moving from the $\gamma$~space to $m$~space, and in particular it translates into a non-decreasing function of~$m$,
as shown by the brown solid line for $B_{\ro{I}} \lesssim 10^{-10}\, \ro{G}$ in \autoref{fig:auger_igmf}.

The translation of the limit becomes more contrived in the presence of  sufficiently strong IGMFs, as is the case of 
$B_{\ro{I}} = 10^{-9}\, \ro{G}$ and $\lambda_{\ro{I}} \gtrsim 10\, \ro{Mpc}$,
which is shown by the brown dashed line.
Along the dashed line where $F > 10^{-19} \, \mathrm{cm^{-2}sr^{-1}s^{-1}}$, 
the MM acceleration is dominated by GMFs and hence
it overlaps with the solid line.
On the other hand at $F \lesssim 10^{-19} \, \mathrm{cm^{-2}sr^{-1}s^{-1}}$, 
the MM acceleration is dominated by IGMFs.
In this regime, the MMs significantly backreact to the IGMFs and thus their velocity also depends on the MM flux 
as $\gamma \propto 1/( m F )$, cf. \autoref{eq:gamma_v_max}.\footnote{With smaller coherence lengths, i.e. $B_{\ro{I}} = 10^{-9}\, \ro{G}$ and $\lambda_{\ro{I}} < 10\, \ro{Mpc}$, the IGMF-induced velocity along (part of) the Auger limit becomes $v_0$ given in \autoref{eq:nonhomoGen}, which  yields a different mass dependence.}
Consequently, a segment of the flux limit with a power-law dependence on the Lorentz factor, 
$F \propto \gamma^{-\alpha}$, translates into 
$F \propto m^{\alpha/(1 - \alpha)}$.
Note that the power $\alpha/(1 - \alpha)$ is negative for $\alpha > 1$, which is the case for $\gamma < 10^9$ 
as one sees in \autoref{fig:limits_exp}.
This explains why the dashed line is negatively tilted around where it merges with the solid.
The dashed line at even smaller flux values has non-negative tilts because the values of $\alpha$ there are smaller than unity.
It should be understood that the parameter space on the left side of the dashed line is excluded. 

\bibliographystyle{unsrtnat}
\bibliography{biblio}

\providecommand{\noopsort}[1]{}\providecommand{\singleletter}[1]{#1}%
\begin{thebibliography}{79}
\providecommand{\natexlab}[1]{#1}
\providecommand{\url}[1]{\texttt{#1}}
\expandafter\ifx\csname urlstyle\endcsname\relax
  \providecommand{\doi}[1]{doi: #1}\else
  \providecommand{\doi}{doi: \begingroup \urlstyle{rm}\Url}\fi

\bibitem['t~Hooft(1974)]{tHooft:1974kcl}
Gerard 't~Hooft.
\newblock {Magnetic Monopoles in Unified Gauge Theories}.
\newblock \emph{Nucl. Phys. B}, 79:\penalty0 276--284, 1974.
\newblock \doi{10.1016/0550-3213(74)90486-6}.

\bibitem[Polyakov(1974)]{Polyakov:1974ek}
Alexander~M. Polyakov.
\newblock {Particle Spectrum in Quantum Field Theory}.
\newblock \emph{JETP Lett.}, 20:\penalty0 194--195, 1974.

\bibitem[Preskill(1979)]{Preskill:1979zi}
John Preskill.
\newblock {Cosmological Production of Superheavy Magnetic Monopoles}.
\newblock \emph{Phys. Rev. Lett.}, 43:\penalty0 1365, 1979.
\newblock \doi{10.1103/PhysRevLett.43.1365}.

\bibitem[Zeldovich and Khlopov(1978)]{Zeldovich:1978wj}
Ya.~B. Zeldovich and M.~Yu. Khlopov.
\newblock {On the Concentration of Relic Magnetic Monopoles in the Universe}.
\newblock \emph{Phys. Lett. B}, 79:\penalty0 239--241, 1978.
\newblock \doi{10.1016/0370-2693(78)90232-0}.

\bibitem[Turner(1982)]{Turner:1982kh}
Michael~S. Turner.
\newblock {Thermal Production of Superheavy Magnetic Monopoles in the Early Universe}.
\newblock \emph{Phys. Lett. B}, 115:\penalty0 95--98, 1982.
\newblock \doi{10.1016/0370-2693(82)90803-6}.

\bibitem[Kobayashi(2021)]{Kobayashi:2021des}
Takeshi Kobayashi.
\newblock {Monopole-antimonopole pair production in primordial magnetic fields}.
\newblock \emph{Phys. Rev. D}, 104\penalty0 (4):\penalty0 043501, 2021.
\newblock \doi{10.1103/PhysRevD.104.043501}.

\bibitem[Dirac(1931)]{Dirac:1931kp}
Paul Adrien~Maurice Dirac.
\newblock {Quantised singularities in the electromagnetic field,}.
\newblock \emph{Proc. Roy. Soc. Lond. A}, 133\penalty0 (821):\penalty0 60--72, 1931.
\newblock \doi{10.1098/rspa.1931.0130}.

\bibitem[Lazarides et~al.(2024)Lazarides, Maji, and Shafi]{Lazarides:2024niy}
George Lazarides, Rinku Maji, and Qaisar Shafi.
\newblock {Quantum tunneling in the early universe: stable magnetic monopoles from metastable cosmic strings}.
\newblock \emph{JCAP}, 05:\penalty0 128, 2024.
\newblock \doi{10.1088/1475-7516/2024/05/128}.

\bibitem[Kephart and Shafi(2025)]{Kephart:2025tik}
Thomas~W. Kephart and Qaisar Shafi.
\newblock {Magnetic Monopoles and Exotic States in $SU(4)_c \times SU(2)_L \times SU(2)_R$}.
\newblock \emph{arXiv preprint}, 3 2025.

\bibitem[Wick et~al.(2003)Wick, Kephart, Weiler, and Biermann]{Wick:2000yc}
Stuart~D. Wick, Thomas~W. Kephart, Thomas~J. Weiler, and Peter~L. Biermann.
\newblock {Signatures for a cosmic flux of magnetic monopoles}.
\newblock \emph{Astropart. Phys.}, 18:\penalty0 663--687, 2003.
\newblock \doi{10.1016/S0927-6505(02)00200-1}.

\bibitem[Perri et~al.(2024)Perri, Bondarenko, Doro, and Kobayashi]{Perri:2023ncd}
Daniele Perri, Kyrilo Bondarenko, Michele Doro, and Takeshi Kobayashi.
\newblock {Monopole acceleration in intergalactic magnetic fields}.
\newblock \emph{Phys. Dark Univ.}, 46:\penalty0 101704, 2024.
\newblock \doi{10.1016/j.dark.2024.101704}.

\bibitem[Giacomelli et~al.(2000)Giacomelli, Giorgini, Lari, Ouchrif, Patrizii, Popa, Spada, and Togo]{Giacomelli:2000de}
G.~Giacomelli, M.~Giorgini, T.~Lari, M.~Ouchrif, L.~Patrizii, V.~Popa, P.~Spada, and V.~Togo.
\newblock {Magnetic monopole bibliography}.
\newblock 5 2000.

\bibitem[Balestra et~al.(2011)Balestra, Giacomelli, Giorgini, Patrizii, Popa, Sahnoun, and Togo]{Balestra:2011ks}
S.~Balestra, G.~Giacomelli, M.~Giorgini, L.~Patrizii, V.~Popa, Z.~Sahnoun, and V.~Togo.
\newblock {Magnetic Monopole Bibliography-II}.
\newblock 5 2011.

\bibitem[Mavromatos and Mitsou(2020)]{Mavromatos:2020gwk}
Nick~E. Mavromatos and Vasiliki~A. Mitsou.
\newblock {Magnetic monopoles revisited: Models and searches at colliders and in the Cosmos}.
\newblock \emph{Int. J. Mod. Phys. A}, 35\penalty0 (23):\penalty0 2030012, 2020.
\newblock \doi{10.1142/S0217751X20300124}.

\bibitem[Ambrosio et~al.(2002{\natexlab{a}})]{MACRO:2002kki}
M.~Ambrosio et~al.
\newblock {The MACRO detector at Gran Sasso}.
\newblock \emph{Nucl. Instrum. Meth. A}, 486:\penalty0 663--707, 2002{\natexlab{a}}.
\newblock \doi{10.1016/S0168-9002(01)02169-6}.

\bibitem[Ambrosio et~al.(2002{\natexlab{b}})]{MACRO:2002jdv}
M.~Ambrosio et~al.
\newblock {Final results of magnetic monopole searches with the MACRO experiment}.
\newblock \emph{Eur. Phys. J. C}, 25:\penalty0 511--522, 2002{\natexlab{b}}.
\newblock \doi{10.1140/epjc/s2002-01046-9}.

\bibitem[Abbasi et~al.(2013)]{IceCube:2012}
R.~Abbasi et~al.
\newblock {Search for Relativistic Magnetic Monopoles with IceCube}.
\newblock \emph{Phys. Rev. D}, 87\penalty0 (2):\penalty0 022001, 2013.
\newblock \doi{10.1103/PhysRevD.87.022001}.

\bibitem[Aartsen et~al.(2014)]{IceCube:2014xnp}
M.~G. Aartsen et~al.
\newblock {Search for non-relativistic Magnetic Monopoles with IceCube}.
\newblock \emph{Eur. Phys. J. C}, 74\penalty0 (7):\penalty0 2938, 2014.
\newblock \doi{10.1140/epjc/s10052-014-2938-8}.
\newblock [Erratum: Eur.Phys.J.C 79, 124 (2019)].

\bibitem[Aartsen et~al.(2016)]{IceCube:2015agw}
M.~G. Aartsen et~al.
\newblock {Searches for Relativistic Magnetic Monopoles in IceCube}.
\newblock \emph{Eur. Phys. J. C}, 76\penalty0 (3):\penalty0 133, 2016.
\newblock \doi{10.1140/epjc/s10052-016-3953-8}.

\bibitem[Abbasi et~al.(2022)]{IceCube:2021eye}
R.~Abbasi et~al.
\newblock {Search for Relativistic Magnetic Monopoles with Eight Years of IceCube Data}.
\newblock \emph{Phys. Rev. Lett.}, 128\penalty0 (5):\penalty0 051101, 2022.
\newblock \doi{10.1103/PhysRevLett.128.051101}.

\bibitem[Aab et~al.(2016)]{PierreAuger:2016imq}
Alexander Aab et~al.
\newblock {Search for ultrarelativistic magnetic monopoles with the Pierre Auger Observatory}.
\newblock \emph{Phys. Rev. D}, 94\penalty0 (8):\penalty0 082002, 2016.
\newblock \doi{10.1103/PhysRevD.94.082002}.

\bibitem[Acharya et~al.(2018)]{CTAConsortium:2017dvg}
B.~S. Acharya et~al.
\newblock {Science with the Cherenkov Telescope Array}.
\newblock 11 2018.
\newblock \doi{10.1142/10986}.

\bibitem[Parker(1970)]{Parker:1970xv}
Eugene~N. Parker.
\newblock {The Origin of Magnetic Fields}.
\newblock \emph{Astrophys. J.}, 160:\penalty0 383, 1970.
\newblock \doi{10.1086/150442}.

\bibitem[Turner et~al.(1982)Turner, Parker, and Bogdan]{Turner:1982ag}
Michael~S. Turner, Eugene~N. Parker, and T.~J. Bogdan.
\newblock {Magnetic Monopoles and the Survival of Galactic Magnetic Fields}.
\newblock \emph{Phys. Rev. D}, 26:\penalty0 1296, 1982.
\newblock \doi{10.1103/PhysRevD.26.1296}.

\bibitem[Widrow(2002)]{Widrow:2002ud}
Lawrence~M. Widrow.
\newblock {Origin of galactic and extragalactic magnetic fields}.
\newblock \emph{Rev. Mod. Phys.}, 74:\penalty0 775--823, 2002.
\newblock \doi{10.1103/RevModPhys.74.775}.

\bibitem[Haverkorn(2014)]{Haverkorn:2014jka}
Marijke Haverkorn.
\newblock {Magnetic Fields in the Milky Way}.
\newblock 6 2014.
\newblock \doi{10.1007/978-3-662-44625-6_17}.

\bibitem[Unger and Farrar(2023)]{Unger:2023lob}
Michael Unger and Glennys~R. Farrar.
\newblock {The Coherent Magnetic Field of the Milky Way}.
\newblock 11 2023.

\bibitem[Durrer and Neronov(2013)]{Durrer:2013pga}
Ruth Durrer and Andrii Neronov.
\newblock {Cosmological Magnetic Fields: Their Generation, Evolution and Observation}.
\newblock \emph{Astron. Astrophys. Rev.}, 21:\penalty0 62, 2013.
\newblock \doi{10.1007/s00159-013-0062-7}.

\bibitem[Alves~Batista and Saveliev(2021)]{AlvesBatista:2021sln}
Rafael Alves~Batista and Andrey Saveliev.
\newblock {The Gamma-ray Window to Intergalactic Magnetism}.
\newblock \emph{Universe}, 7\penalty0 (7):\penalty0 223, 2021.
\newblock \doi{10.3390/universe7070223}.

\bibitem[Neronov et~al.(2021)Neronov, Semikoz, and Kalashev]{Neronov:2021xua}
Andrii Neronov, Dmitri Semikoz, and Oleg Kalashev.
\newblock {Limit on intergalactic magnetic field from ultra-high-energy cosmic ray hotspot in Perseus-Pisces region}.
\newblock 12 2021.

\bibitem[Kobayashi and Perri(2023)]{Kobayashi:2023ryr}
Takeshi Kobayashi and Daniele Perri.
\newblock {Parker bounds on monopoles with arbitrary charge from galactic and primordial magnetic fields}.
\newblock \emph{Phys. Rev. D}, 108\penalty0 (8):\penalty0 083005, 2023.
\newblock \doi{10.1103/PhysRevD.108.083005}.

\bibitem[Tavecchio et~al.(2010)Tavecchio, Ghisellini, Foschini, Bonnoli, Ghirlanda, and Coppi]{Tavecchio:2010mk}
F.~Tavecchio, G.~Ghisellini, L.~Foschini, G.~Bonnoli, G.~Ghirlanda, and P.~Coppi.
\newblock {The intergalactic magnetic field constrained by Fermi/LAT observations of the TeV blazar 1ES 0229+200}.
\newblock \emph{Mon. Not. Roy. Astron. Soc.}, 406:\penalty0 L70--L74, 2010.
\newblock \doi{10.1111/j.1745-3933.2010.00884.x}.

\bibitem[{Neronov} and {Vovk}(2010)]{Neronov2010}
Andrii {Neronov} and Ievgen {Vovk}.
\newblock {Evidence for Strong Extragalactic Magnetic Fields from Fermi Observations of TeV Blazars}.
\newblock \emph{Science}, 328\penalty0 (5974):\penalty0 73, April 2010.
\newblock \doi{10.1126/science.1184192}.

\bibitem[{Ackermann} et~al.(2018)]{Ackermann2018}
M.~{Ackermann} et~al.
\newblock {The Search for Spatial Extension in High-latitude Sources Detected by the Fermi Large Area Telescope}.
\newblock \emph{Astroph. J. Suppl.}, 237\penalty0 (2):\penalty0 32, August 2018.
\newblock \doi{10.3847/1538-4365/aacdf7}.

\bibitem[Acciari et~al.(2023)]{MAGIC:2022piy}
V.~A. Acciari et~al.
\newblock {A lower bound on intergalactic magnetic fields from time variability of 1ES 0229+200 from MAGIC and Fermi/LAT observations}.
\newblock \emph{Astron. Astrophys.}, 670:\penalty0 A145, 2023.
\newblock \doi{10.1051/0004-6361/202244126}.

\bibitem[Aharonian et~al.(2023)]{HESS:2023zwb}
F.~Aharonian et~al.
\newblock {Constraints on the Intergalactic Magnetic Field Using Fermi-LAT and H.E.S.S. Blazar Observations}.
\newblock \emph{Astrophys. J. Lett.}, 950\penalty0 (2):\penalty0 L16, 2023.
\newblock \doi{10.3847/2041-8213/acd777}.

\bibitem[Blunier et~al.(2025)Blunier, Neronov, and Semikoz]{Blunier:2025ddu}
J.~Blunier, A.~Neronov, and D.~Semikoz.
\newblock {Revision of conservative lower bound on intergalactic magnetic field from Fermi and Cherenkov telescope observations of extreme blazars}.
\newblock \emph{arXiv preprint}, 6 2025.

\bibitem[Barrow et~al.(1997)Barrow, Ferreira, and Silk]{Barrow:1997}
John~D. Barrow, Pedro~G. Ferreira, and Joseph Silk.
\newblock Constraints on a primordial magnetic field.
\newblock \emph{Phys. Rev. Lett.}, 78:\penalty0 3610--3613, May 1997.
\newblock \doi{10.1103/PhysRevLett.78.3610}.
\newblock URL \url{https://link.aps.org/doi/10.1103/PhysRevLett.78.3610}.

\bibitem[Ade et~al.(2016)]{Planck:2015zrl}
P.~A.~R. Ade et~al.
\newblock {Planck 2015 results. XIX. Constraints on primordial magnetic fields}.
\newblock \emph{Astron. Astrophys.}, 594:\penalty0 A19, 2016.
\newblock \doi{10.1051/0004-6361/201525821}.

\bibitem[Jedamzik and Saveliev(2019)]{Jedamzik:2018itu}
Karsten Jedamzik and Andrey Saveliev.
\newblock {Stringent Limit on Primordial Magnetic Fields from the Cosmic Microwave Background Radiation}.
\newblock \emph{Phys. Rev. Lett.}, 123\penalty0 (2):\penalty0 021301, 2019.
\newblock \doi{10.1103/PhysRevLett.123.021301}.

\bibitem[Pavi\v{c}evi\'c et~al.(2025)Pavi\v{c}evi\'c, Ir\v{s}i\v{c}, Viel, Bolton, Haehnelt, Martin-Alvarez, Puchwein, and Ralegankar]{Pavicevic:2025gqi}
Mak Pavi\v{c}evi\'c, Vid Ir\v{s}i\v{c}, Matteo Viel, James Bolton, Martin~G. Haehnelt, Sergio Martin-Alvarez, Ewald Puchwein, and Pranjal Ralegankar.
\newblock {Constraints on Primordial Magnetic Fields from the Lyman-$\alpha$ forest}.
\newblock 1 2025.

\bibitem[Pignataro et~al.(2025)Pignataro, O'Sullivan, Bonafede, Bernardi, Vazza, and Carretti]{Pignataro:2025ntd}
G.~V. Pignataro, S.~P. O'Sullivan, A.~Bonafede, G.~Bernardi, F.~Vazza, and E.~Carretti.
\newblock {Detection of magnetic fields in superclusters of galaxies}.
\newblock 3 2025.

\bibitem[Jedamzik et~al.(2025)Jedamzik, Pogosian, and Abel]{Jedamzik:2025cax}
Karsten Jedamzik, Levon Pogosian, and Tom Abel.
\newblock {Hints of Primordial Magnetic Fields at Recombination and Implications for the Hubble Tension}.
\newblock 3 2025.

\bibitem[Grasso and Rubinstein(2001)]{Grasso:2000wj}
Dario Grasso and Hector~R. Rubinstein.
\newblock {Magnetic fields in the early universe}.
\newblock \emph{Phys. Rept.}, 348:\penalty0 163--266, 2001.
\newblock \doi{10.1016/S0370-1573(00)00110-1}.

\bibitem[Parker(1987)]{Parker:1987}
Eugene~N. Parker.
\newblock {Magnetic Monopole Plasma Oscillations and the Survival of Galactic Magnetic Fields}.
\newblock \emph{Astrophys. J.}, 321:\penalty0 349, 1987.
\newblock \doi{10.1086/165633}.

\bibitem[Kogut et~al.(1993)]{Kogut:1993ag}
A.~Kogut et~al.
\newblock {Dipole anisotropy in the COBE DMR first year sky maps}.
\newblock \emph{Astrophys. J.}, 419:\penalty0 1, 1993.
\newblock \doi{10.1086/173453}.

\bibitem[Abbasi et~al.(2023)]{TelescopeArray:2023sbd}
R.~U. Abbasi et~al.
\newblock {An extremely energetic cosmic ray observed by a surface detector array}.
\newblock \emph{Science}, 382\penalty0 (6673), 2023.
\newblock \doi{10.1126/science.abo5095}.

\bibitem[Hogan et~al.(2008)Hogan, Besson, Ralston, Kravchenko, and Seckel]{Hogan:2008sx}
D.~P. Hogan, D.~Z. Besson, J.~P. Ralston, I.~Kravchenko, and D.~Seckel.
\newblock {Relativistic Magnetic Monopole Flux Constraints from RICE}.
\newblock \emph{Phys. Rev. D}, 78:\penalty0 075031, 2008.
\newblock \doi{10.1103/PhysRevD.78.075031}.

\bibitem[{Spengler} and {Schwanke}(2011)]{Spengler:2011}
G.~{Spengler} and U.~{Schwanke}.
\newblock Signatures of ultrarelativistic magnetic monopoles in imaging atmorpheric cherenkov telescopes.
\newblock In \emph{Proceedings to the 32nd ICRC}, October 2011.

\bibitem[Antipin et~al.(2008)]{BAIKAL:2007kno}
K.~Antipin et~al.
\newblock {Search for relativistic magnetic monopoles with the Baikal Neutrino Telescope}.
\newblock \emph{Astropart. Phys.}, 29:\penalty0 366--372, 2008.
\newblock \doi{10.1016/j.astropartphys.2008.03.006}.

\bibitem[Detrixhe et~al.(2011)Detrixhe, Besson, Gorham, Allison, Baughmann, Beatty, Belov, Bevan, Binns, Chen, Chen, Clem, Connolly, De~Marco, Dowkontt, DuVernois, Frankenfeld, Grashorn, Hogan, Griffith, Hill, Hoover, Israel, Javaid, Liewer, Matsuno, Mercurio, Miki, Mottram, Nam, Nichol, Palladino, Romero-Wolf, Ruckman, Saltzberg, Seckel, Varner, Vieregg, and Wang]{Detrixhe:2011dan}
M.~Detrixhe, D.~Besson, P.~W. Gorham, P.~Allison, B.~Baughmann, J.~J. Beatty, K.~Belov, S.~Bevan, W.~R. Binns, C.~Chen, P.~Chen, J.~M. Clem, A.~Connolly, D.~De~Marco, P.~F. Dowkontt, M.~A. DuVernois, C.~Frankenfeld, E.~W. Grashorn, D.~P. Hogan, N.~Griffith, B.~Hill, S.~Hoover, M.~H. Israel, A.~Javaid, K.~M. Liewer, S.~Matsuno, B.~C. Mercurio, C.~Miki, M.~Mottram, J.~Nam, R.~J. Nichol, K.~Palladino, A.~Romero-Wolf, L.~Ruckman, D.~Saltzberg, D.~Seckel, G.~S. Varner, A.~G. Vieregg, and Y.~Wang.
\newblock Ultrarelativistic magnetic monopole search with the anita-ii balloon-borne radio interferometer.
\newblock \emph{Physical Review D}, 83\penalty0 (2), 2011.
\newblock ISSN 1550-2368.
\newblock \doi{10.1103/physrevd.83.023513}.
\newblock URL \url{http://dx.doi.org/10.1103/PhysRevD.83.023513}.

\bibitem[Albert et~al.(2025)]{ANTARES:2025ojl}
A.~Albert et~al.
\newblock {Search for Magnetic Monopoles with the Complete ANTARES Dataset}.
\newblock \emph{arXiv preprint}, 5 2025.

\bibitem[Ahlen(1978)]{Ahlen:1978}
S.~P. Ahlen.
\newblock Stopping-power formula for magnetic monopoles.
\newblock \emph{Phys. Rev. D}, 17:\penalty0 229--233, Jan 1978.
\newblock \doi{10.1103/PhysRevD.17.229}.
\newblock URL \url{https://link.aps.org/doi/10.1103/PhysRevD.17.229}.

\bibitem[Ahlen(1980)]{RevModPhys.52.121}
Steven~P. Ahlen.
\newblock Theoretical and experimental aspects of the energy loss of relativistic heavily ionizing particles.
\newblock \emph{Rev. Mod. Phys.}, 52:\penalty0 121--173, Jan 1980.
\newblock \doi{10.1103/RevModPhys.52.121}.
\newblock URL \url{https://link.aps.org/doi/10.1103/RevModPhys.52.121}.

\bibitem[Ahlen and Kinoshita(1982)]{Ahlen:1982}
S.~P. Ahlen and K.~Kinoshita.
\newblock Calculation of the stopping power of very-low-velocity magnetic monopoles.
\newblock \emph{Phys. Rev. D}, 26:\penalty0 2347--2363, Nov 1982.
\newblock \doi{10.1103/PhysRevD.26.2347}.
\newblock URL \url{https://link.aps.org/doi/10.1103/PhysRevD.26.2347}.

\bibitem[Tompkins(1965)]{Tompkins:1965}
Donald~R. Tompkins.
\newblock Total energy loss and \ifmmode \check{C}\else \v{C}\fi{}erenkov emission from monopoles.
\newblock \emph{Phys. Rev.}, 138:\penalty0 B248--B250, Apr 1965.
\newblock \doi{10.1103/PhysRev.138.B248}.
\newblock URL \url{https://link.aps.org/doi/10.1103/PhysRev.138.B248}.

\bibitem[Christy(2011)]{Christy:2011lza}
Brian~John Christy.
\newblock \emph{A Search for Relativistic Magnetic Monopoles with the IceCube 22-String Detector}.
\newblock PhD thesis, Maryland U., College Park, 2011.

\bibitem[{Spengler}(2009)]{Spengler:2009}
G~{Spengler}.
\newblock {Signatures of Ultrarelativistic Magnetic Monopoles in Imaging Cherenkov Telescopes}.
\newblock Master's thesis, Humboldt-University of Berlin, 2009.

\bibitem[Oceanic et~al.(1976)Oceanic, Administration, Aeronautics, Administration, and Force]{USStandardAtmosphere:1976dan}
National Oceanic, Atmospheric Administration, National Aeronautics, Space Administration, and United States~Air Force.
\newblock \emph{U.S. Standard Atmosphere, 1976}.
\newblock U.S. Government Printing Office, Washington, D.C., 1976.
\newblock URL \url{https://ntrs.nasa.gov/citations/19770009539}.
\newblock NOAA-S/T 76-1562.

\bibitem[Donath et~al.(2023)]{Gammapy:2023gvb}
Axel Donath et~al.
\newblock {Gammapy: A Python package for gamma-ray astronomy}.
\newblock \emph{Astron. Astrophys.}, 678:\penalty0 A157, 2023.
\newblock \doi{10.1051/0004-6361/202346488}.

\bibitem[Schwinger(1951)]{Schwinger:1951nm}
Julian~S. Schwinger.
\newblock {On gauge invariance and vacuum polarization}.
\newblock \emph{Phys. Rev.}, 82:\penalty0 664--679, 1951.
\newblock \doi{10.1103/PhysRev.82.664}.

\bibitem[Affleck and Manton(1982)]{Affleck:1981ag}
Ian~K. Affleck and Nicholas~S. Manton.
\newblock {Monopole Pair Production in a Magnetic Field}.
\newblock \emph{Nucl. Phys. B}, 194:\penalty0 38--64, 1982.
\newblock \doi{10.1016/0550-3213(82)90511-9}.

\bibitem[Affleck et~al.(1982)Affleck, Alvarez, and Manton]{Affleck:1981bma}
Ian~K. Affleck, Orlando Alvarez, and Nicholas~S. Manton.
\newblock {Pair Production at Strong Coupling in Weak External Fields}.
\newblock \emph{Nucl. Phys. B}, 197:\penalty0 509--519, 1982.
\newblock \doi{10.1016/0550-3213(82)90455-2}.

\bibitem[Gould and Rajantie(2017)]{Gould:2017zwi}
Oliver Gould and Arttu Rajantie.
\newblock {Magnetic monopole mass bounds from heavy ion collisions and neutron stars}.
\newblock \emph{Phys. Rev. Lett.}, 119\penalty0 (24):\penalty0 241601, 2017.
\newblock \doi{10.1103/PhysRevLett.119.241601}.

\bibitem[Gould et~al.(2019)Gould, Ho, and Rajantie]{Gould:2019myj}
Oliver Gould, David L.~J. Ho, and Arttu Rajantie.
\newblock {Towards Schwinger production of magnetic monopoles in heavy-ion collisions}.
\newblock \emph{Phys. Rev. D}, 100\penalty0 (1):\penalty0 015041, 2019.
\newblock \doi{10.1103/PhysRevD.100.015041}.

\bibitem[Tumasyan et~al.(2022)]{CMS:2021rwb}
Armen Tumasyan et~al.
\newblock {Search for strongly interacting massive particles generating trackless jets in proton\textendash{}proton collisions at $\sqrt{s} = 13\,\text {TeV} $}.
\newblock \emph{Eur. Phys. J. C}, 82\penalty0 (3):\penalty0 213, 2022.
\newblock \doi{10.1140/epjc/s10052-022-10095-5}.

\bibitem[Aad et~al.(2023)]{ATLAS:2023esy}
Georges Aad et~al.
\newblock {Search for magnetic monopoles and stable particles with high electric charges in $ \sqrt{s} $ = 13 TeV pp collisions with the ATLAS detector}.
\newblock \emph{JHEP}, 11:\penalty0 112, 2023.
\newblock \doi{10.1007/JHEP11(2023)112}.

\bibitem[Acharya et~al.(2022{\natexlab{a}})]{MoEDAL:2021mpi}
B.~Acharya et~al.
\newblock {Search for highly-ionizing particles in $pp$ collisions at the LHC\textquoteright{}s Run-1 using the prototype MoEDAL detector}.
\newblock \emph{Eur. Phys. J. C}, 82\penalty0 (8):\penalty0 694, 2022{\natexlab{a}}.
\newblock \doi{10.1140/epjc/s10052-022-10608-2}.

\bibitem[Pinfold et~al.(2009)]{MoEDAL:2009jwa}
James Pinfold et~al.
\newblock {Technical Design Report of the MoEDAL Experiment}.
\newblock 6 2009.

\bibitem[Acharya et~al.(2022{\natexlab{b}})]{MoEDAL:2021vix}
B.~Acharya et~al.
\newblock {Search for magnetic monopoles produced via the Schwinger mechanism}.
\newblock \emph{Nature}, 602\penalty0 (7895):\penalty0 63--67, 2022{\natexlab{b}}.
\newblock \doi{10.1038/s41586-021-04298-1}.

\bibitem[Aad et~al.(2025)]{ATLAS:2024nzp}
Georges Aad et~al.
\newblock {Search for Magnetic Monopole Pair Production in Ultraperipheral Pb+Pb Collisions at sNN=5.36\,\,TeV with the ATLAS Detector at the LHC}.
\newblock \emph{Phys. Rev. Lett.}, 134\penalty0 (6):\penalty0 061803, 2025.
\newblock \doi{10.1103/PhysRevLett.134.061803}.

\bibitem[Hook and Huang(2017)]{Hook:2017vyc}
Anson Hook and Junwu Huang.
\newblock {Bounding millimagnetically charged particles with magnetars}.
\newblock \emph{Phys. Rev. D}, 96\penalty0 (5):\penalty0 055010, 2017.
\newblock \doi{10.1103/PhysRevD.96.055010}.

\bibitem[Kobayashi and Perri(2022)]{Kobayashi:2022qpl}
Takeshi Kobayashi and Daniele Perri.
\newblock {Parker bound and monopole pair production from primordial magnetic fields}.
\newblock \emph{Phys. Rev. D}, 106\penalty0 (6):\penalty0 063016, 2022.
\newblock \doi{10.1103/PhysRevD.106.063016}.

\bibitem[Adams et~al.(1993)Adams, Fatuzzo, Freese, Tarle, Watkins, and Turner]{Adams:1993fj}
Fred~C. Adams, Marco Fatuzzo, Katherine Freese, Gregory Tarle, Richard Watkins, and Michael~S. Turner.
\newblock {Extension of the Parker bound on the flux of magnetic monopoles}.
\newblock \emph{Phys. Rev. Lett.}, 70:\penalty0 2511--2514, 1993.
\newblock \doi{10.1103/PhysRevLett.70.2511}.

\bibitem[Rephaeli and Turner(1983)]{Rephaeli:1982nv}
Yoel Rephaeli and Michael~S. Turner.
\newblock {The Magnetic Monopole Flux and the Survival of Intracluster Magnetic Fields}.
\newblock \emph{Phys. Lett. B}, 121:\penalty0 115--118, 1983.
\newblock \doi{10.1016/0370-2693(83)90897-3}.

\bibitem[Long and Vachaspati(2015)]{Long:2015cza}
Andrew~J. Long and Tanmay Vachaspati.
\newblock {Implications of a Primordial Magnetic Field for Magnetic Monopoles, Axions, and Dirac Neutrinos}.
\newblock \emph{Phys. Rev. D}, 91:\penalty0 103522, 2015.
\newblock \doi{10.1103/PhysRevD.91.103522}.

\bibitem[Aghanim et~al.(2020)]{Planck:2018vyg}
N.~Aghanim et~al.
\newblock {Planck 2018 results. VI. Cosmological parameters}.
\newblock \emph{Astron. Astrophys.}, 641:\penalty0 A6, 2020.
\newblock \doi{10.1051/0004-6361/201833910}.
\newblock [Erratum: Astron.Astrophys. 652, C4 (2021)].

\bibitem[Mitri(1996)]{demitri}
{Ivan}~De Mitri.
\newblock \emph{Ricerca di Monopoli Magnetici supermassivi nella radiazione cosmica}.
\newblock PhD thesis, Università degli Studi di L'Aquila, 1996.

\bibitem[Derkaoui et~al.(1998)Derkaoui, Giacomelli, Lari, Margiotta, Ouchrif, Patrizii, Popa, and Togo]{Derkaoui:1998uv}
J.~Derkaoui, G.~Giacomelli, T.~Lari, A.~Margiotta, M.~Ouchrif, L.~Patrizii, V.~Popa, and V.~Togo.
\newblock {Energy losses of magnetic monopoles and of dyons in the earth}.
\newblock \emph{Astropart. Phys.}, 9:\penalty0 173--183, 1998.
\newblock \doi{10.1016/S0927-6505(98)00016-4}.

\end{thebibliography}

\end{document}